\documentclass[10pt]{article}
\usepackage[letterpaper,margin=1in]{geometry}
\usepackage[T1]{fontenc}
\usepackage{lmodern}
\usepackage{microtype}
\AddToHook{env/verbatim/begin}{\microtypesetup{protrusion=false}}
\usepackage{amsmath,amssymb,amsthm}
\usepackage{booktabs,tabularx,multirow}
\usepackage[table]{xcolor}
\usepackage{tikz}
\usetikzlibrary{arrows.meta,positioning,fit,calc,backgrounds,shapes.geometric}
\usepackage{pgfplots}
\pgfplotsset{compat=1.18}
\usepackage[numbers,sort&compress]{natbib}
\usepackage{caption}
\usepackage{enumitem}
\usepackage[colorlinks=true,linkcolor=blue!50!black,citecolor=blue!50!black,urlcolor=blue!50!black]{hyperref}
\hypersetup{pdftitle={FCP: Formal Context Protocol for Verifying LLM-Generated Artifacts},pdfauthor={Shashank}}

\setlist{itemsep=1pt,topsep=3pt}
\makeatletter
\renewcommand{\bibfont}{\small\@clubpenalty\@M}
\makeatother
\newtheorem{definition}{Definition}
\newtheorem{proposition}{Proposition}
\newtheorem{corollary}{Corollary}

\newcommand{\sem}[1]{[\![#1]\!]}
\newcommand{\ok}{\mathit{ok}}
\newcommand{\FCPPL}{FCP-PL}
\newcommand{\code}[1]{\texttt{#1}}
\newcommand{\RQ}[1]{\textbf{RQ#1}}

\definecolor{youc}{RGB}{43,68,198}
\definecolor{okc}{RGB}{28,122,75}
\definecolor{badc}{RGB}{182,50,42}
\definecolor{askc}{RGB}{143,92,0}

\title{\textbf{FCP: Formal Context Protocol}\\[3pt]
\textbf{for Verifying LLM-Generated Artifacts}}
\author{Shashank\\[2pt]\small\href{mailto:sshashan@alumni.usc.edu}{\texttt{sshashan@alumni.usc.edu}}}
\date{November 2026}

\begin{document}
\maketitle

\begin{abstract}
Language models now write hardware designs, prompt responses, tool-using agents and code, together with the tests and assertions meant to check them. When the same kind of system writes an artifact and its checks, a passing check is weak evidence: a single misreading of the request yields both a faulty artifact and a check that accepts it. We present the Formal Context Protocol (FCP), which checks AI-generated artifacts against the request itself, independently of their own tests and assertions. FCP separates \emph{conformance} (does the artifact satisfy a formal reading of a requirement?), which model checkers and SMT solvers decide, from \emph{fidelity} (does the reading mean what the prose means?), which admits no decision procedure and is refutable by a concrete witness. Each of eight ways in which a proof can certify a requirement falsely becomes a mechanical check that returns a witness, and what no machine can settle becomes a list of concrete questions for the request's author. Adapters place four kinds of artifact on the same engines: hardware directly, text as a labeled trace, an agent as a circuit whose model output is unconstrained, and code, which is run on every input of a bounded scope drawn from the request and, where it translates, proved for all inputs. On four hardware designs, trusting AI-written assertions certified 23\% of violated requirements, and FCP certified none. On ten code tasks, code written by three language models passed its own tests (18 to 745 per suite) and violated the request in every case, and in nine tasks all three models resolved the planted conflict identically. Trusting the tests certified all 33 violations. FCP found nine of the ten planted conflicts before any code existed and certified three violations, all on one task, where both formal readings and all three models resolved the same ambiguity alike. Across these studies, agreement among models, readings or tests did not imply correctness.
\end{abstract}

\section{Introduction}
\label{sec:intro}

A developer who asks an AI system for a design, an answer or a program receives more output than line-by-line review can cover, and each established safeguard has a known limitation. Tests and assertions written by the same system are circular: one misreading produces both the wrong artifact and the check that accepts it, and language models do not reliably correct their own reasoning without external feedback~\cite{huang2024selfcorrect}. An LLM judge returns a single verdict, can be manipulated by the text it evaluates~\cite{raina2024judgerobust}, and favors its own generations~\cite{panickssery2024selfpref}. Formal verification removes the sampling limit of testing and makes the property the critical object: a property that misstates the requirement, once proved, gives false assurance for every behavior.

\begin{figure}[t]
\centering
\begin{tikzpicture}
\pgfplotsset{resolution/.style={scale only axis, width=0.265\linewidth, height=3.3cm,
  xmin=43, xmax=57, ymin=46.2, ymax=60.2, clip=false,
  axis lines=left, axis line style={gray!75, -}, tick style={gray!75}, tick align=outside,
  ytick={50,54.98}, yticklabels={}, ymajorgrids, grid style={gray!28, very thin},
  tick label style={font=\scriptsize}, label style={font=\small},
  title style={font=\scriptsize, align=center, yshift=-3pt}},
  written/.style={gray!60, line width=0.5pt}, charge/.style={line width=1.25pt, line join=round}}
\newcommand{\aswritten}{\addplot[written] coordinates {(43,47.99) (49.99,54.98)};
  \addplot[written] coordinates {(50,50) (57,57)};}
\begin{axis}[resolution, name=sonnet, xtick={50}, xticklabels={\$50.00},
  yticklabels={\$50.00,\$54.98}, ylabel={amount charged},
  title={{\small\textbf{Sonnet} and the reference}\\keep the drop}]
\draw[gray!28, very thin] (axis cs:50,46.2) -- (axis cs:50,50);
\addplot[charge, badc] coordinates {(43,47.99) (49.99,54.98)};
\addplot[charge, badc] coordinates {(50,50) (57,57)};
\draw[badc, line width=0.6pt, densely dashed] (axis cs:50,54.98) -- (axis cs:50,50);
\addplot[only marks, mark=*, mark size=1.7pt, badc, mark options={fill=white}] coordinates {(49.99,54.98)};
\addplot[only marks, mark=*, mark size=1.7pt, badc] coordinates {(50,50)};
\node[font=\scriptsize, anchor=north west, align=left] at (axis cs:43.3,60.0) {breaks rule 7:\\\$4.98 less for\\one cent more};
\end{axis}
\begin{axis}[resolution, name=opus, at={(sonnet.east)}, anchor=west, xshift=0.045\linewidth,
  xtick={45.01,50}, xticklabels={\$45.01,\$50.00},
  xlabel={cart subtotal, no coupon, not a member},
  title={{\small\textbf{Opus}}\\holds the charge at \$50.00}]
\fill[youc, opacity=0.13] (axis cs:45.01,50) -- (axis cs:49.99,54.98) -- (axis cs:50,50) -- cycle;
\draw[gray!28, very thin] (axis cs:45.01,46.2) -- (axis cs:45.01,50) (axis cs:50,46.2) -- (axis cs:50,50);
\aswritten
\addplot[charge, youc] coordinates {(43,47.99) (45.01,50) (50,50) (57,57)};
\node[font=\scriptsize, anchor=north west, align=left] at (axis cs:43.3,60.0) {breaks rule 6:\\the shipping fee\\shrinks to zero};
\end{axis}
\begin{axis}[resolution, name=haiku, at={(opus.east)}, anchor=west, xshift=0.045\linewidth,
  xtick={50,54.98}, xticklabels={\$50.00,\$54.98},
  title={{\small\textbf{Haiku}}\\holds the charge at \$54.98}]
\fill[askc, opacity=0.15] (axis cs:49.99,54.98) -- (axis cs:54.98,54.98) -- (axis cs:50,50) -- cycle;
\draw[gray!28, very thin] (axis cs:50,46.2) -- (axis cs:50,50) (axis cs:54.98,46.2) -- (axis cs:54.98,54.98);
\aswritten
\addplot[charge, askc] coordinates {(43,47.99) (49.99,54.98) (54.98,54.98) (57,57)};
\node[font=\scriptsize, anchor=north west, align=left] at (axis cs:43.3,60.0) {breaks rule 6:\\a fee is still\\charged from \$50};
\end{axis}
\end{tikzpicture}
\caption{Three resolutions of an unsatisfiable prompt (\S\ref{sec:eval-code}). Rule 6 (free shipping from \$50) and rule 7 (adding items never lowers the amount charged) cannot both hold: as written, the charge falls from \$54.98 to \$50.00 between subtotals of \$49.99 and \$50.00 (thin gray line). Shading marks the subtotals on which a model departs from the rules as written. FCP finds the calls at 4,999 and 5,000 cents before any code exists. All three models identified the conflict and passed their own tests, and each violated a rule.}
\label{fig:conflict}
\end{figure}
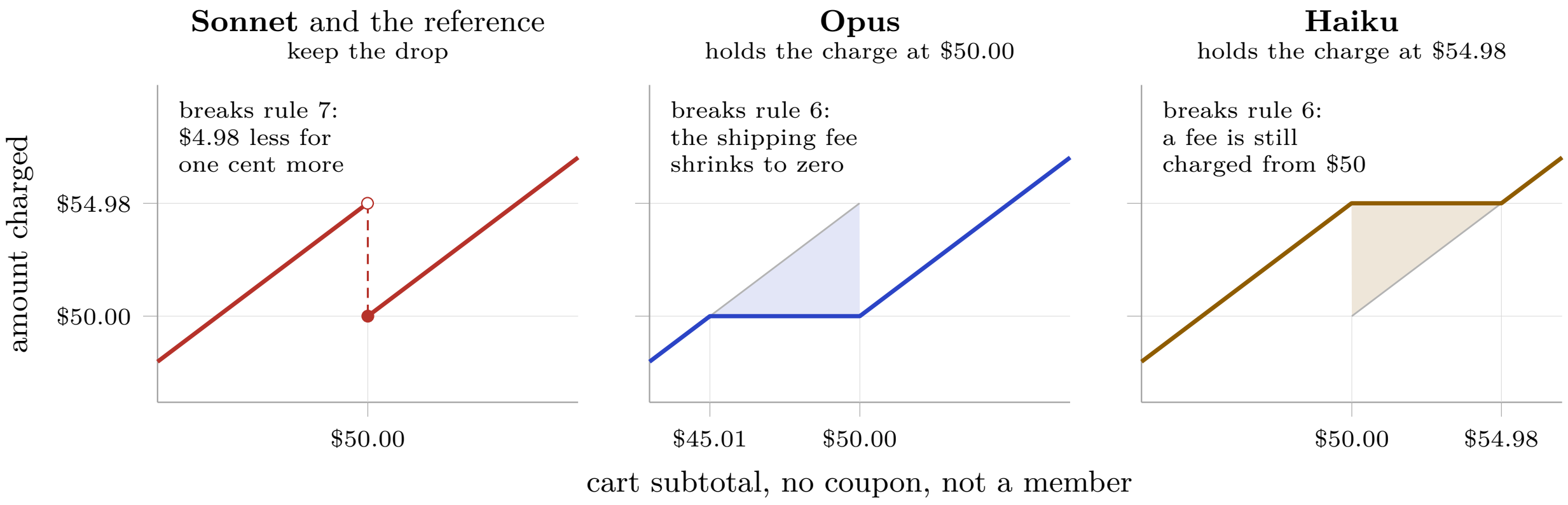

Figure~\ref{fig:conflict} gives a small example. A store asks for a function \code{checkout(subtotal, coupon, member)} with seven rules, two of which cannot both hold: shipping is free from \$50, and adding items to the cart must never lower the amount charged. A \$49.99 cart costs \$54.98 with shipping, and a \$50.00 cart costs \$50.00. We gave the prompt to three language models. Each model identified the conflict, resolved it differently, wrote between 134 and 745 tests, and passed all of them. Each implementation violates a rule of the prompt, and a flow that certifies a requirement whenever the model's own tests pass certifies all three violations.

This paper presents the Formal Context Protocol (FCP). Its goal is to let a person establish whether an AI-generated artifact meets the request, and where it falls short, without reading the artifact. FCP checks the artifact against the request: it proves what holds, replays every failure on the artifact itself, and reduces the rest to concrete questions that only the request's author can answer. On the checkout prompt, FCP finds the conflicting pair of calls before any code exists, certifies none of the three violations, and asks three questions: which rule yields at the threshold, which amount the \$50 refers to, and how half a cent is rounded.

The central difficulty is that the meaning of prose cannot be checked mechanically. FCP therefore separates the two questions on which every verification claim rests. \emph{Conformance} asks whether the artifact satisfies a formal reading of a requirement; standard engines decide it. \emph{Fidelity} asks whether the reading means what the prose means; no procedure decides it, and it is refutable: one concrete behavior on which two independent readings disagree refutes at least one, and their agreement corroborates both. We identify eight situations in which a proof certifies a requirement falsely and give each a mechanical witness (Table~\ref{tab:modes}). Only fidelity ends in a question, phrased as an example: a waveform, a conversation or a function call.

\paragraph{Contributions.}
\begin{itemize}
\item A decomposition of false assurance into eight situations, each with a mechanical witness, and a protocol that produces these witnesses: a provenance-anchored intermediate representation, isolated independent formalizers, and verdicts that carry witnesses (\S\ref{sec:setting} and \S\ref{sec:protocol}).
\item Analyses with precise semantics: conflicts as unrealizability witnesses, forcing stimuli, determinacy, and comparison of readings as assume-guarantee contracts in the context of the other requirements. In the hardware study, contracts are needed for soundness and context raises acceptance 4.4-fold (\S\ref{sec:analyses}).
\item Adapters that place four kinds of artifact on the same engines (\S\ref{sec:adapters}). Text becomes a labeled trace. An agent is checked with its model as a cut point, which assigns every rule a trust class: \emph{enforced}, \emph{delegated} or \emph{broken}. Code is run on an exhaustive scope derived from the request, for functions, strings and lists, objects, changes to existing code, code that acts through services and files, and records. For functions, comparing two readings reduces to a bounded number of calls (Proposition~\ref{prop:horizon}).
\item Proofs beyond the scope: translation-validated symbolic execution deciding code requirements for every input, every script of service answers, and call sequences of every length (\S\ref{sec:proofs}).
\item An evaluation on four hardware designs, a prompt, an agent and ten code tasks, in which the formalizers, RTL generators, agents and code writers are language models (\S\ref{sec:eval}). Self-check baselines certified 23\% of violated hardware requirements and all 33 violations in code. FCP certified none of the hardware violations and 3 of the 33 code violations, which an additional reading of the ambiguous phrase removes. Beyond the scope, 2,245 of 2,263 verdicts are decided for every input and agree with the scope; eleven hold only up to a bound, four of which fail beyond it, and seven fail at inputs the scope missed.
\end{itemize}

One finding recurs across our studies: each form of agreement we measured (code with its author's tests, three models' code on millions of calls, two isolated formalizers, two labelers) held at least once on an incorrect result. FCP is therefore designed around disagreement, which it turns into a concrete witness (\S\ref{sec:discussion}).

\section{Problem Setting}
\label{sec:setting}

\paragraph{Traces, properties and contracts.} An interface $\Sigma=(I,O)$ consists of input and output signals. A trace is a sequence $\pi:\mathbb{N}\to \mathit{Val}(I\cup O)$ in which reset is asserted at cycle~$0$. An artifact $M$ denotes a set of traces $\sem{M}$. A safety property $\varphi$ is recognized by a deterministic monitor~\cite{kupferman2001safety} with a verdict $\ok^\varphi_t(\pi)$ at each cycle $t$, and $\pi\models\varphi$ iff $\ok^\varphi_t(\pi)$ holds for all $t$. The \emph{spec-only model} $\mathcal{S}_\Sigma$ contains every trace over $\Sigma$.

\begin{definition}[Assume-guarantee satisfaction]
\label{def:ag}
For assumptions $A$ and a property $\varphi$, $M\models(A\Rightarrow\varphi)$ iff for every $\pi\in\sem{M}$ and every $t$, if $\ok^A_{t'}(\pi)$ holds for all $t'\le t$, then $\ok^\varphi_t(\pi)$ holds.
\end{definition}

\paragraph{Readings and false assurance.} A request is decomposed into atomic requirements. For a requirement $r$, a \emph{reading} is a pair $(A_r,\varphi_r)$ of assumptions and a property, written by a \emph{formalizer}: a person or a language model. The contract $C^r=(A_r\Rightarrow\varphi_r)$ denotes the set of traces that satisfy it in the sense of Definition~\ref{def:ag}.

\begin{definition}[False assurance]
\label{def:fa}
Let $(A^\star_r,\varphi^\star_r)$ be a reading that captures the author's intent, the \emph{gold reading}. A flow gives \emph{false assurance} on $(M,r)$ if it certifies $r$ for $M$ and $M\not\models(A^\star_r\Rightarrow\varphi^\star_r)$.
\end{definition}

The gold reading is used only in the evaluation; our gold readings are validated mechanically.

\paragraph{Conformance and fidelity.} Verification decides \emph{conformance}, $M\models(A_r\Rightarrow\varphi_r)$. A proof of conformance is evidence for $r$ only if the reading is \emph{faithful} to the prose, and it carries no evidence in any situation of Table~\ref{tab:modes}. Every situation other than F concerns objects that FCP holds (the artifact, the readings and the interface), so each has a mechanical witness. Because $r$ is prose, fidelity admits no decision procedure, and it is refutable. If two independently written readings disagree, one behavior accepted by one reading and rejected by the other shows that at least one of them is unfaithful, and the witness a person then judges is a six-cycle waveform or a two-call example. Agreement between independent readings corroborates both, in the tradition of N-version programming~\cite{avizienis1985nversion}. The known caveat of that tradition also applies: independently produced versions fail together more often than independence would predict~\cite{knight1986independence}. Our studies measure this effect for language models (\S\ref{sec:eval-agree}).

\begin{table}[t]
\caption{Situations in which a proof certifies a requirement falsely, and the witness FCP returns for each. Fidelity (F) is the one situation without a mechanical decision; its witness is a concrete behavior for a person to judge.}
\label{tab:modes}
\centering\small
\begin{tabularx}{\linewidth}{@{}l>{\raggedright\arraybackslash}X>{\raggedright\arraybackslash}X@{}}
\toprule
& Situation in which a proof of $M\models(A\Rightarrow\varphi)$ certifies $r$ falsely & Mechanical witness returned by FCP \\
\midrule
V & $\varphi$ never triggers on $M$ under $A$, or holds on every behavior that $A$ allows & proof that the trigger is unreachable; proof of $\varphi$ on $\mathcal{S}_\Sigma$ (tautology) \\
E & $A$ excludes behaviors of $M$ itself (assumptions may restrict only the environment) & reachable state with no legal non-reset input (non-receptiveness) \\
F & $(A,\varphi)$ differs from the meaning of $r$ & behavior accepted by one independent reading and rejected by another, in the context of the other requirements \\
C & the requirements contradict each other & input sequence on which no output sequence meets every guarantee (unrealizability) \\
U & the request does not determine some output & two conforming behaviors with equal history that differ in an output \\
W & the properties are too weak to notice a change of $M$ & observable mutant of $M$ that no property kills \\
D & a rule holds only because a black-box component, such as an agent's model, behaves & violation with the component's output unconstrained that disappears under its contract \\
T & an engine or a model of the artifact is wrong & disagreement between engines (PDR and BMC); mismatch on replay in a simulator \\
\bottomrule
\end{tabularx}
\end{table}

\section{The Formal Context Protocol}
\label{sec:protocol}

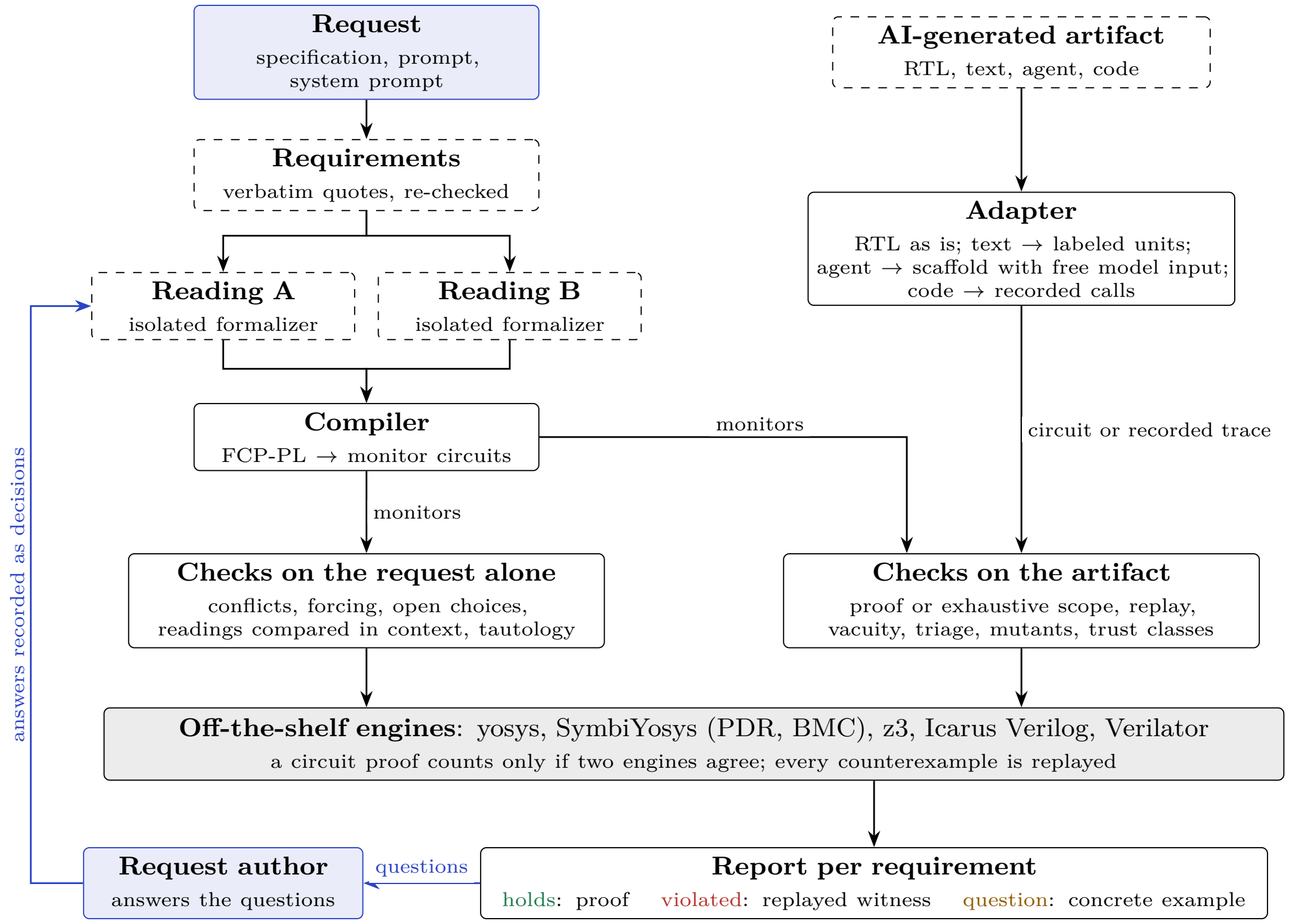
\begin{figure}[t]
\centering
\resizebox{0.86\linewidth}{!}{%
\begin{tikzpicture}[font=\small,
  box/.style={draw, rounded corners=2pt, align=center, inner sep=3pt, text width=#1, minimum height=8mm},
  box/.default=33mm,
  ai/.style={box=#1, dashed}, ai/.default=33mm,
  you/.style={box=#1, fill=youc!10, draw=youc}, you/.default=33mm,
  eng/.style={box=#1, fill=gray!15}, eng/.default=33mm,
  arr/.style={-{Stealth[length=2mm]}, semithick},
  lab/.style={font=\scriptsize, fill=white, inner sep=2pt}]
\node[you=40mm] (req) at (0,0) {\textbf{Request}\\[3pt]\scriptsize specification, prompt,\\ system prompt\par};
\node[ai=40mm] (rq) at (0,-1.5) {\textbf{Requirements}\\\scriptsize verbatim quotes, re-checked};
\node[ai=30mm] (ra) at (-1.75,-3.1) {\textbf{Reading A}\\\scriptsize isolated formalizer};
\node[ai=30mm] (rb) at (1.75,-3.1) {\textbf{Reading B}\\\scriptsize isolated formalizer};
\node[box=40mm] (cp) at (0,-4.7) {\textbf{Compiler}\\\scriptsize \FCPPL{} $\to$ monitor circuits};
\node[ai=44mm] (ans) at (8,0) {\textbf{AI-generated artifact}\\\scriptsize RTL, text, agent, code};
\node[box=50mm] (ad) at (8,-2.4) {\textbf{Adapter}\\[3pt]\scriptsize RTL as is; text $\to$ labeled units;\\ agent $\to$ scaffold with free model input;\\ code $\to$ recorded calls\par};
\node[box=56mm] (c1) at (0,-6.7) {\textbf{Checks on the request alone}\\[3pt]\scriptsize conflicts, forcing, open choices,\\ readings compared in context, tautology\par};
\node[box=56mm] (c2) at (8,-6.7) {\textbf{Checks on the artifact}\\[3pt]\scriptsize proof or exhaustive scope, replay,\\ vacuity, triage, mutants, trust classes\par};
\node[eng=142mm] (en) at (4,-8.45) {\textbf{Off-the-shelf engines}: yosys, SymbiYosys (PDR, BMC), z3, Icarus Verilog, Verilator\\\scriptsize a circuit proof counts only if two engines agree; every counterexample is replayed};
\node[box=94mm] (rp) at (6.2,-10.15) {\textbf{Report per requirement}\\\scriptsize \textcolor{okc}{holds}: proof \quad \textcolor{badc}{violated}: replayed witness \quad \textcolor{askc}{question}: concrete example};
\node[you=32mm] (hu) at (-1.75,-10.15) {\textbf{Request author}\\\scriptsize answers the questions};
\draw[arr] (req) -- (rq);
\draw[arr] (rq.south) -- ++(0,-0.3) -| (ra.north);
\draw[arr] (rq.south) -- ++(0,-0.3) -| (rb.north);
\draw[arr] (ra.south) -- ++(0,-0.35) -| (cp.north);
\draw[arr] (rb.south) -- ++(0,-0.35) -| (cp.north);
\draw[arr] (ans) -- (ad);
\draw[arr] (cp) -- node[lab, right]{monitors} (c1);
\draw[arr] (cp.east) -| node[lab, pos=0.3, above]{monitors} ($(c2.north)+(-1.4,0)$);
\draw[arr] (ad) -- node[lab, right]{circuit or recorded trace} (c2);
\draw[arr] (c1) -- (c1|-en.north);
\draw[arr] (c2) -- (c2|-en.north);
\draw[arr] (en.south -| rp.north) -- (rp.north);
\draw[arr, youc] (rp) -- node[lab, above]{questions} (hu);
\draw[arr, youc] (hu.west) -- (-4.1,-10.15) -- node[lab, pos=0.5, sloped, above]{answers recorded as decisions} (-4.1,-3.1) -- (ra.west);
\end{tikzpicture}}
\caption{The FCP process. Dashed boxes are written by AI systems and are untrusted; solid boxes are deterministic code; shaded boxes are off-the-shelf engines. Formalizers are not given the artifact or each other's work. The checks on the left need no artifact, so a request is analyzed before any artifact exists.}
\label{fig:pipeline}
\end{figure}

Figure~\ref{fig:pipeline} shows the process: language models decompose and formalize the request, and deterministic code compiles the readings, runs the analyses and reports each requirement as holding, violated or a question.

\paragraph{Intermediate representation.} The IR has three layers, each written in YAML and validated by a JSON Schema. The \emph{design manifest} declares the interface, environment defaults, a \emph{frame} (properties that every behavior satisfies by construction of the encoding), and one of: RTL variants, a cut point (\S\ref{sec:agents}) or a function (\S\ref{sec:code}). \emph{Requirements} are atomic; each carries a verbatim quote, a kind (guarantee, assumption, scenario, definition or non-functional) and any recorded ambiguities. Line spans are computed from the quotes and re-verified on every load, so provenance cannot be fabricated. \emph{Readings} are sets of properties, assumptions, scenarios and don't-cares written in \FCPPL{}, with ghost state, rigid (universally quantified) variables and named definitions. A requirement left unformalized must state why.

\paragraph{\FCPPL{}.} The property language is a past-time temporal logic over Booleans, bit-vectors and integers with statically computed ranges (Appendix~\ref{app:lang}). Arithmetic is exact: every subexpression is compiled to a two's-complement wire wide enough for its interval, so $\mathit{count}-1$ evaluates to $-1$ at zero, with no wrap-around. Logical operators require one-bit operands. Future obligations exist only inside patterns (\code{next}, \code{within}, \code{hold\_until}, \code{bounded\_wait}, \code{inorder}), whose monitors flag the minimal bad prefix. A \code{prev} that reaches before the current reset segment disables evaluation, so that no arbitrary state is read, and a trigger that looks back $n$ cycles cannot arm an obligation in the first $n$ cycles. Every property is compiled twice: an assert form, which may use nondeterminism (for instance, Wolper's data-independence argument~\cite{wolper1986data} for \code{inorder}), and a deterministic form for every analysis that assumes the property.

\paragraph{Design rules.} Four rules shape the protocol.
(R1)~\emph{Separation of roles.} Language models write only requirements, readings, labels and adapters. A deterministic compiler and standard engines decide everything else, and a circuit proof counts only if two engines agree.
(R2)~\emph{Isolation.} Formalizers are not given the artifact or each other's work.
(R3)~\emph{Black-box properties.} Readings refer only to interface signals and ghost state, so a reading applies to any implementation and can be analyzed before one exists.
(R4)~\emph{Witness-carrying verdicts.} Every verdict other than a proof carries a concrete trace that a person can judge.{\widowpenalty=10000\par}

\paragraph{Verdicts.} Each requirement receives a level stating what has been established (Table~\ref{tab:verdicts}). It is ASSURED only if it is proved, its trigger is witnessed, the environment is receptive, and an independent reading is equivalent to it \emph{as a contract}. Every other requirement enters a review queue with its witness.

\begin{table}[t]
\caption{Verdict levels. For code, a requirement is \emph{certified} if it holds on the exhaustive scope under both independent readings, the readings are equivalent in context, it lies in no conflict, and its text carries no ambiguity flag; it is \emph{proved} if it also holds for all inputs, or PROVED-BOUNDED if for all call sequences up to a bound (\S\ref{sec:proofs}).}
\label{tab:verdicts}
\centering\small
\begin{tabularx}{\linewidth}{@{}l>{\raggedright\arraybackslash}X@{}}
\toprule
Level & Meaning and attached evidence \\
\midrule
ASSURED & proved by two engines; trigger witnessed; assumptions receptive; an independent reading is an equivalent contract (in context, \S\ref{sec:analyses}) \\
PROVED-DISPUTED & proved, and an independent reading differs; the simplest distinguishing behavior is attached \\
PROVED-CONFLICTED & proved, and the requirement belongs to a minimal conflict; the conflicting stimulus is attached \\
VACUOUS & trigger unreachable under the assumptions (proof attached), or the property is a tautology \\
ENV-UNSOUND & the assumptions are not receptive; a blocking state is attached \\
FAILED & counterexample replayed in simulation and triaged to an artifact deviation or a conflict \\
BOUNDED, UNKNOWN & the engines did not conclude; never counted as a proof \\
UNFORMALIZED & no property; the formalizer's reason is recorded \\
\bottomrule
\end{tabularx}
\end{table}

\section{Analyses}
\label{sec:analyses}

All analyses are generated harnesses around the same compiled monitors, run by SymbiYosys~\cite{symbiyosys} with ABC PDR~\cite{bradley2011ic3,een2011pdr} and Boolector BMC~\cite{brummayer2009boolector,biere1999bmc}, or by z3~\cite{demoura2008z3} on the SMT2 model from yosys~\cite{yosys}.

\paragraph{Verification and replay.} PDR and BMC run independently; a PDR pass together with a BMC failure is an engine disagreement and does not count as a proof. Every counterexample is replayed in Icarus Verilog~\cite{icarus} from its witness and compared cycle by cycle.

\paragraph{Vacuity and tautology.} For a property with a trigger, a cover on $M$ under $A$ yields either a witness or a PDR proof that the trigger is unreachable, in which case the property is VACUOUS~\cite{beer1997vacuity,kupferman2003vacuity}. Independently, $\mathcal{S}_\Sigma\models(A\Rightarrow\varphi)$ means that $\varphi$ constrains nothing.

\paragraph{Receptiveness.} An assumption is legitimate only if it restricts the environment. A \emph{blocking state} is a reachable state $s$ such that $\neg A(s,i)$ for every $i\in\mathit{Val}(I)$ with $i(\mathit{reset})=0$. Finding one is an exists-forall query, which FCP solves by counterexample-guided refinement.

\begin{definition}[Conflict]
\label{def:conflict}
A finite input sequence $\sigma\in\mathit{Val}(I)^{k+1}$ is a \emph{conflict} of a set of readings if, for every output sequence $\tau$, some guarantee fails on $(\sigma,\tau)$ at a cycle $t\le k$ up to which every assumption held.
\end{definition}

A conflict is an unrealizability witness~\cite{pnueli1989synthesis}: an environment that plays $\sigma$ defeats every implementation (Proposition~\ref{prop:conflict}). A simpler candidate definition, ``a trigger situation that cannot occur under all the requirements'', also flags situations that a specification legitimately excludes; on a consistent bridge specification, it reported dozens of spurious conflicts. FCP draws candidates from every trigger situation, every pair of requirements and every scenario, and extends each by a lookahead $L=1+w$, where $w$ is the longest obligation window; the lookahead is required because a prefix that shows no violation so far may already admit no conforming continuation. A minimized unsatisfiable core names the minimal set of conflicting requirements. \emph{Forcing} searches for $\sigma$ directly, by exists-forall counterexample-guided synthesis over output sequences. Both searches are sound for the conflicts they report and incomplete against adaptive environments.

\begin{proposition}[Reported conflicts are sound]
\label{prop:conflict}
If $\sigma$ is a conflict of a set of readings, every artifact that accepts $\sigma$ violates one of their guarantees on $\sigma$, at a cycle up to which all assumptions have held.
\end{proposition}

\paragraph{Determinacy.} By self-composition~\cite{barthe2004selfcomposition,clarkson2010hyperproperties}, FCP searches for traces $\pi,\pi'$ that both conform, agree on all signals before cycle $t$ and on the inputs at $t$, and differ at $t$ in an output that is not a declared don't-care. Such a pair is a concrete instance of under-specification.

\begin{definition}[Contract comparison in context]
\label{def:compare}
Let $C_x^r=(A_x\Rightarrow\varphi^r_x)$ be the contract of reading $x$ for requirement $r$. Reading $a$ implies $b$ on $r$ iff $C_a^r\subseteq C_b^r$ over $\mathcal{S}_\Sigma$, where witnesses must be extendable by $L$ cycles. \emph{In context}, both sides are intersected with $K^r=\bigcap_{r'\neq r}(C^{r'}_a\cap C^{r'}_b)$, the behaviors that meet every other requirement under both readings. The relation is EQUIVALENT, STRONGER, WEAKER or INCOMPARABLE, and each failed direction yields the simplest distinguishing behavior.
\end{definition}

Soundness requires comparing contracts, each property under its own reading's assumptions. In an earlier version, which pooled the assumptions, one formalizer's masking assumption entered the agreement check and caused the only false assurances that version gave (\S\ref{sec:eval-hw}). Comparing in context removes disputes that arise only on behaviors that other requirements already exclude.

\paragraph{Triage and mutation.} For a counterexample, FCP asks whether any conforming output sequence exists for the same inputs. If so, MaxSAT finds the closest and localizes the defect to a (cycle, signal, expected, actual) difference. If none exists, even under assume-guarantee semantics, the requirements conflict on that stimulus and an unsatisfiable core names them. Mutation coverage~\cite{chockler2003coverage,jia2011mutation} checks every property on every observable mutant of $M$; a surviving mutant marks, at a source line, behavior that no requirement constrains.

\section{Adapters: Every Artifact as a Circuit}
\label{sec:adapters}

The engines check circuits; an adapter for each kind of artifact determines how exact the verdict is.

\subsection{Text responses}
A response is split into units (sentences or lines), one per cycle. Each unit receives labels (atomic propositions such as ``states a feature''), and the prompt becomes a small automaton over them. The \emph{factorization hypothesis} is that the author's intent depends on the response only through its label trace. Labels are assigned by two isolated labelers and form the trust boundary: a verdict is exact relative to its labels, and disagreements between labelers are surfaced like disagreements between readings. The analyses also test the label semantics: the first consistency check produced a ``conforming'' email that hid benefit statements inside closing units, which no real text does. A frame rule now forbids any other label on a closing unit.

\subsection{Agents: the model as a cut point}
\label{sec:agents}
\begin{figure}[t]
\centering
\begin{tikzpicture}[font=\small,
  box/.style={draw, rounded corners=2pt, align=center, inner sep=3pt, text width=#1, minimum height=9mm}, box/.default=30mm,
  arr/.style={-{Stealth[length=2mm]}, semithick}, lab/.style={font=\scriptsize, fill=white, inner sep=2pt}]
\node[box=30mm, dashed] (m) at (0,0) {\textbf{Model}\\\scriptsize cut out of the proof};
\node[box=28mm] (cu) at (-4.9,-2.0) {\textbf{Customer}\\\scriptsize messages, buttons};
\node[box=34mm, very thick] (sc) at (0,-2.0) {\textbf{Scaffold} (the circuit)\\\scriptsize Python $\to$ Verilog};
\node[box=28mm] (to) at (4.9,-2.0) {\textbf{Tools}\\[3pt]\scriptsize verify, lookup,\\ refund, \dots\par};
\node[box=34mm] (mo) at (0,-3.9) {\textbf{Monitors of the prompt}\\\scriptsize flag each violating turn};
\draw[arr, very thick] (m) -- node[lab, right, align=left]{proposed call $X$:\\ free, any value} (sc);
\draw[arr] (cu) -- node[lab, above]{buttons} (sc);
\draw[arr] ($(sc.east)+(0,0.15)$) -- node[lab, above]{what runs} ($(to.west)+(0,0.15)$);
\draw[arr] ($(to.west)+(0,-0.15)$) -- node[lab, below]{result} ($(sc.east)+(0,-0.15)$);
\draw[arr, dashed] (cu.north) |- node[lab, pos=0.75, above]{messages} (m.west);
\draw[arr, dashed] (to.north) |- node[lab, pos=0.75, above]{results, notes} (m.east);
\draw[arr] (sc) -- (mo);
\node[align=left, font=\scriptsize, anchor=west] at (2.4,-3.9) {\textcolor{okc}{$\bullet$} \textbf{enforced}: no proposal violates it\\ \textcolor{askc}{$\bullet$} \textbf{delegated}: rests on the model\\ \textcolor{badc}{$\bullet$} \textbf{broken}: the scaffold violates it};
\end{tikzpicture}
\caption{An agent checked as a circuit. One turn is one cycle. An enforced rule holds for every call the model could propose, on any input the model reads, including text planted in a tool result.}
\label{fig:agent}
\end{figure}
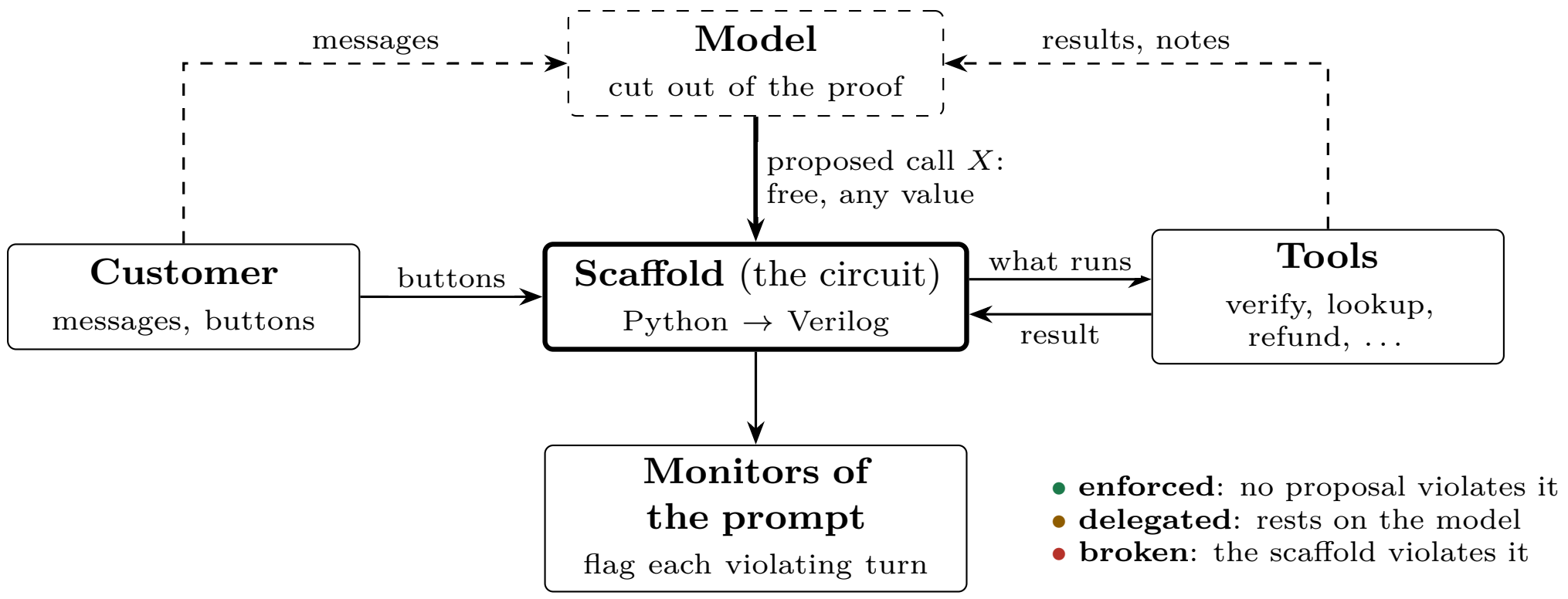

A tool-using agent runs a loop: the model proposes a call, the \emph{scaffold} decides what runs, and the result feeds the next turn. We write the scaffold's decision logic once, in a Python subset, run it in the agent, and translate it to Verilog, checked by co-simulation. The model's proposal $X$ is a free input, so a proof covers every output of the model, including one that follows instructions injected into a tool result~\cite{greshake2023injection} (Figure~\ref{fig:agent}).

\begin{definition}[Trust classes]
\label{def:trust}
Let $\mathit{pass}_o(X)$ be the value that output $o$ takes when the scaffold passes the proposal through unchanged, and let the \emph{model contract} $\hat C_r$ be $\varphi_r$ with every current-cycle output $o$ outside temporal operators replaced by $\mathit{pass}_o(X)$. With $X$ free, $r$ is \emph{enforced} if $M\models(A\Rightarrow\varphi_r)$; \emph{delegated} if $M\not\models(A\Rightarrow\varphi_r)$ and $M\models(A\wedge\hat C_r\Rightarrow\varphi_r)$; and \emph{broken} otherwise.
\end{definition}

\begin{proposition}[Meaning of the trust classes]
\label{prop:trust}
If $r$ is enforced, every run of the agent satisfies $r$ under $A$, for all proposals of the model. If $r$ is delegated, every run satisfies $r$ under $A$ up to the first turn in which the model's proposal violates $\hat C_r$.
\end{proposition}

The delegated rules are the agent's residual trust in its model. A counterexample on a pass-through scaffold identifies the guard that is needed; a counterexample that survives the model contract is a scaffold bug; and recorded runs are checked against the contracts by runtime verification~\cite{leucker2009rv}. The construction is related to shield synthesis~\cite{bloem2015shield}, with the shield written by hand and its guarantees proved, and to designs that confine an agent's model by construction~\cite{debenedetti2025camel}.

\subsection{Code: a function as a circuit without memory}
\label{sec:code}
Code written by an assistant does not, in general, compile to a circuit. FCP therefore treats the two sides separately: the request side stays on the circuit engines, and the code is executed.

\paragraph{Interface.} A function $f:D_1\times\dots\times D_n\to R$ becomes a design with one input signal per argument. Its outputs are the returned values (one signal per tuple element) and an outcome code $\omega$: $0$ if the function returned values that fit their signals; $1,\dots,m$ for the listed exceptions; $m{+}1$ for any other exception, or no answer within the time limit; and $m{+}2$ for any other return value, such as a float where an integer is due. Each cycle is one call, calls are independent, and the checker chooses them and their order. A requirement that relates two calls, such as monotonicity, is a property over consecutive cycles that uses \code{prev}.

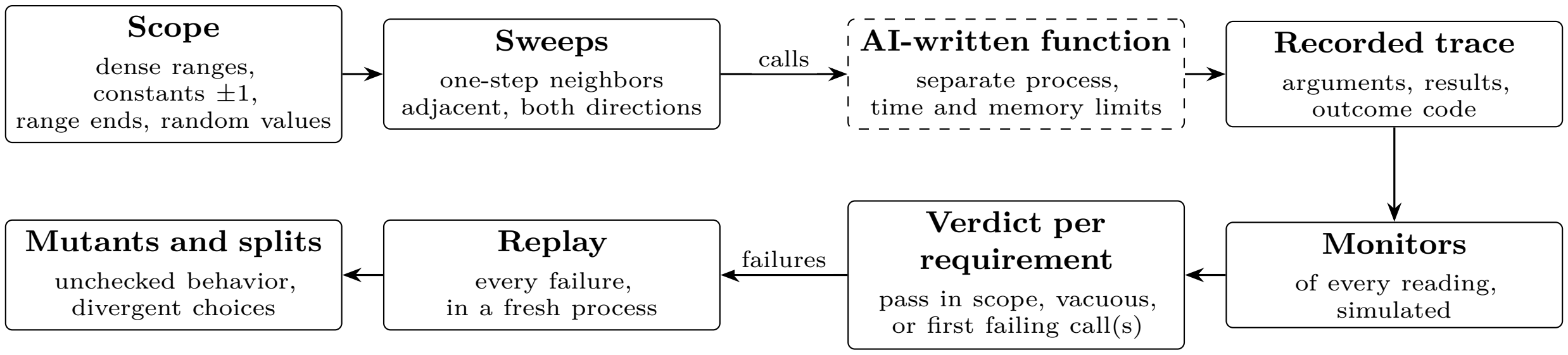
\begin{figure}[t]
\centering
\begin{tikzpicture}[font=\small,
  box/.style={draw, rounded corners=2pt, align=center, inner sep=3pt, text width=#1, minimum height=11mm}, box/.default=33mm,
  arr/.style={-{Stealth[length=2mm]}, semithick}, lab/.style={font=\scriptsize, fill=white, inner sep=2pt}]
\node[box] (sc) at (0,0) {\textbf{Scope}\\[3pt]\scriptsize dense ranges,\\ constants $\pm1$,\\ range ends, random values\par};
\node[box] (sw) at (3.95,0) {\textbf{Sweeps}\\[3pt]\scriptsize one-step neighbors\\ adjacent, both directions\par};
\node[box, dashed] (f) at (8.8,0) {\textbf{AI-written function}\\[3pt]\scriptsize separate process,\\ time and memory limits\par};
\node[box] (tr) at (12.75,0) {\textbf{Recorded trace}\\[3pt]\scriptsize arguments, results,\\ outcome code\par};
\node[box] (mo) at (12.75,-2.1) {\textbf{Monitors}\\[3pt]\scriptsize of every reading,\\ simulated\par};
\node[box] (v) at (8.8,-2.1) {\textbf{Verdict per requirement}\\[3pt]\scriptsize pass in scope, vacuous,\\ or first failing call(s)\par};
\node[box] (rp) at (3.95,-2.1) {\textbf{Replay}\\[3pt]\scriptsize every failure,\\ in a fresh process\par};
\node[box] (mu) at (0,-2.1) {\textbf{Mutants and splits}\\[3pt]\scriptsize unchecked behavior,\\ divergent choices\par};
\draw[arr] (sc) -- (sw); \draw[arr] (sw) -- node[lab, above]{calls} (f); \draw[arr] (f) -- (tr);
\draw[arr] (tr) -- (mo); \draw[arr] (mo) -- (v); \draw[arr] (v) -- node[lab, above]{failures} (rp); \draw[arr] (rp) -- (mu);
\end{tikzpicture}
\caption{The artifact side of the code adapter. The request side (conflicts, open choices and comparison of readings) runs on the readings alone, before any code exists.}
\label{fig:code}
\end{figure}

\paragraph{Scope and evidence.} The code is called on a scope $D_s$ that contains, for each argument, a dense range; every integer constant of the readings, the prompt and the code, together with its neighbors $c-1$ and $c+1$; the ends of the signal's range; and random values. Calls are ordered in \emph{sweeps}, so that calls that differ by one scope step in one argument are adjacent; on a smaller scope the sweeps also run in reverse. The compiled monitors check the recorded trace, and every failure is replayed in a fresh process. The evidence is exhaustive within $D_s$, in the small-scope tradition~\cite{jackson2006alloy}, and makes no claim beyond $D_s$; \S\ref{sec:proofs} closes that gap where the code permits. One-operator mutants of the code are run on the scope, and an observable mutant that changes no verdict marks behavior that no rule checks (Figure~\ref{fig:code}).

\begin{proposition}[Bounded horizon]
\label{prop:horizon}
Let readings $a$ and $b$ of a function consist of \code{implies} and invariant items without assumptions, ghost state or unbounded history operators, each looking back at most $d$ calls. Then $a$ and $b$ disagree on a requirement $r$, plainly or in context, on some trace iff they disagree on a trace of at most $d+1$ calls.
\end{proposition}

An item that relates a call to its successor looks across two calls, and the horizon counts both; an adversarial review of the engine (\S\ref{sec:eval-agree}) found that the implementation had omitted the trigger call of such items, and the corrected horizon moves no stored result. The proof (Appendix~\ref{app:proofs}) rests on two facts: the function has no state, so any window of calls is itself a trace, and a look-back that reaches before reset disables an item, which then holds. The readings of the checkout task satisfy the hypotheses. Proposition~\ref{prop:horizon} reduces comparison of such readings to a single solver query over $d+1$ calls. Bit-level engines time out on this query: two readings that write the same product differently, such as $s\cdot(100-p)$ and $(100-p)\cdot s$ with $p=\min(c+10m,30)$, pose a multiplier-equivalence problem. FCP therefore also encodes \FCPPL{} directly in integer arithmetic, where both expressions become the same polynomial. This encoding is a second implementation of the semantics, so every witness it finds is confirmed on the compiled monitors, and every agreement is spot-checked on the bit-level model.

\paragraph{Boundary conflicts.} Trigger-situation candidates take arbitrary values, and forcing does not converge over wide arithmetic inputs. FCP therefore also builds pairs of calls from the constants of the readings and the prompt: consecutive boundary values of one argument, in both orders, with the other arguments at their constants. For each pair, an incremental solver asks whether any pair of outputs meets every property (Definition~\ref{def:conflict} with $k=2$) and returns a minimal core. Such conflicts are sound by Proposition~\ref{prop:conflict}.

\paragraph{Open choices and code splits.} For each output that determinacy finds open, FCP reports the simplest call on which two values are allowed. When several implementations of one prompt exist, the calls on which they answer differently are grouped by how they split the implementations, a form of differential testing~\cite{mckeeman1998differential}. Each split is a decision that the prompt left to the author of the code.

\subsection{Beyond functions: strings, objects, changes, effects and records}
\label{sec:beyond}
\paragraph{Strings and lists.} A string or list argument is a bounded sequence $s=(\ell,\langle s_0,\dots,s_{n-1}\rangle)$ of at most $n$ slots of $w$ bits, carried by two ports. The monitors read it in canonical form, $s_k:=0$ for $k\ge\ell$, so nothing a port carries beyond the length can change a verdict, and the spec-only harnesses constrain free sequence ports to that form. \FCPPL{} gains $\mathit{len}(s)$, $s[i]$ (zero outside $s$), $x\in s$, equality of sequences, string and list literals, ASCII character tests, and bounded quantifiers $\mathit{all}$, $\mathit{any}$, $\mathit{count}$ and $\mathit{sum}$ over $i<b$, where $b$ has a static maximum $\bar b\le 256$. The compiler expands a quantifier into $\bar b$ guarded terms, for example $\mathit{all}(i<b{:}\ \phi)\equiv\bigwedge_{k<\bar b}(k<b\rightarrow\phi[k/i])$, so a sequence property remains a combinational function of one cycle and every engine applies unchanged. The integer encoding represents $s$ by $\ell,s_0,\dots,s_{n-1}$ with $0\le\ell\le n$ and $k\ge\ell\rightarrow s_k=0$.

\paragraph{A scope drawn from the request.} Strings have too many values for dense ranges, and the relevant boundaries are often properties of the content: a card number is valid from 12 digits, at any length. Let $\mathcal{A}$ contain, for every comparison $x\bowtie y$ of two integers in any formula, the atoms $x-y=\delta$ for $\delta\in\{-1,0,1\}$; for every other condition $c$, both $c$ and $\neg c$; and $\ell=n$ for every length $n$ that the readings name. Let $\mathcal{C}$ contain \emph{true} and the trigger of every rule. For each $(C,\alpha)\in\mathcal{C}\times\mathcal{A}$, z3 solves $C\wedge\alpha$ in the integer encoding with the outputs free and completes the input at random from an alphabet that contains the characters of the readings and the code and one character of each class. Each such call is followed by its neighbors at edit distance one, which cross the boundary on which $\alpha$ lies, and then by all short and some random sequences: constrained-random stimulus, as in hardware verification, with the constraints taken from the request. A rule on a computed value, such as \emph{a subtotal of \$10.00 or more}, yields no input at its boundary when the outputs are free. Every atom that mentions an output is therefore also solved with the outputs obeying one reading (the consequent of each item whose trigger holds), so that the inputs produce exactly the boundary value. A conflict, once found, is shrunk to the simplest stimulus that still conflicts.

\paragraph{Objects.} For a method of a class, reset creates the object and every later cycle calls the method on it. The object's state stays inside the code, like unnamed registers of an RTL design, and readings may keep a specification model in ghost state. The scope is a set of call sequences: for every condition in $\mathcal{A}$, the shortest sequence (found by bounded model checking on $\mathcal{S}_\Sigma$) after which the condition holds, with every rule satisfied up to that point, together with short and random walks and their one-edit neighbors. A rule quantified over a large rigid range, such as \emph{every customer}, has no expansion that can be assumed, so the conflict search assumes it at the values to which a candidate sequence binds the rigid variable, one monitor instance each. Because $\forall u\,\varphi(u)$ implies every pinned instance, a stimulus on which no output meets the pinned rules is a conflict for the full rules (Corollary~\ref{cor:pin}).

\paragraph{Several methods.} A class with several methods is one circuit. An $\mathit{op}$ input selects the method of each call, methods may share an argument signal, arguments that the called method does not take are $0$ in every engine, and each method has its own result signals. The class may belong to a package of several modules.

\paragraph{Changes.} A change request has two specifications: the request itself, and the code before the change for everything the request does not touch. With a \emph{baseline}, the old code runs on the same calls and its answers drive additional input signals, so a rule can relate the new answer to the old one, as in $\neg\mathit{student}\rightarrow\mathit{price}=\mathit{old\_price}\wedge\dots$. A regression is then a call on which no rule permits the difference. On the code side and in the conflict search, the old answers are those of the old code; in the spec-only analyses they are free, so readings are compared for any old code.

\paragraph{Effects and records.} Code that acts through a service, received as an argument or imported by name, is given a recording stand-in. Every call it makes is an output event, every answer is a scripted input, and a run longer than a bound ends as ``did not finish'', so that a rule such as ``keep trying until it succeeds'' can fail. Files the code opens become stand-in handles whose methods are events, with a per-run virtual disk. A record is one signal per field, and a list of records is one sequence per field, the fields forming a group of equal length in every engine; the runner builds the dictionaries or dataclasses and decomposes the result.

\paragraph{Engines.} Unrolled quantifiers yield monitors of tens of thousands of lines, so traces are replayed on a cached Verilator build of each module, which agreed with Icarus Verilog on every check we compared and is about $400\times$ faster on string monitors. A comparison that the integer encoding cannot decide within a minute is checked call by call on the scope and then on the bit-level model; a comparison that both leave open stays UNKNOWN and does not count as agreement.

\section{Proofs Beyond the Scope}
\label{sec:proofs}

\begin{figure}[t]
\centering
\begin{tikzpicture}[font=\small,
  box/.style={draw, rounded corners=2pt, align=center, inner sep=3pt, text width=#1, minimum height=9mm}, box/.default=30mm,
  arr/.style={-{Stealth[length=2mm]}, semithick}, lab/.style={font=\scriptsize, fill=white, inner sep=2pt},
  note/.style={font=\scriptsize, text=black!65, align=center}]
\node[box, dashed] (code) at (0,0) {\textbf{AI-written code}\\\scriptsize Python source};
\node[box] (sym) at (4.1,0) {\textbf{Symbolic execution}\\\scriptsize terms $T(f)$ per path};
\node[box] (rd) at (0,-2.2) {\textbf{Reading}\\\scriptsize \FCPPL{} items};
\node[box] (enc) at (4.1,-2.2) {\textbf{Integer encoding}\\\scriptsize $E$ over $k$ calls};
\node[note, above=2pt of sym] {validated: equals the code\\ on every call of the scope};
\node[note, below=2pt of enc] {validated: fails exactly where\\ the monitors fail on the scope};
\node[box=34mm] (q) at (8.6,-1.1) {\textbf{z3}: do some $k$ calls\\ in the interface's ranges\\ violate $r$?};
\node[box=24mm, draw=okc, text=okc] (pr) at (12.6,-0.2) {\textbf{PROVED}};
\node[box=24mm, draw=badc, text=badc] (fa) at (12.6,-2.0) {\textbf{FAILED}\\[3pt]\scriptsize replayed on code\\ and monitors\par};
\draw[arr] (code) -- (sym); \draw[arr] (rd) -- (enc);
\draw[arr] (sym) -- (q); \draw[arr] (enc) -- (q);
\draw[arr] (q) -- node[lab, above, sloped]{unsat} (pr);
\draw[arr] (q) -- node[lab, below, sloped]{sat} (fa);
\end{tikzpicture}
\caption{Proofs beyond the scope. The translation $T(f)$ and the encoding $E$ are second semantics: each must reproduce the code or the monitors on every call of the scope before z3 is asked, and a mismatch blocks the proof.}
\label{fig:prove}
\end{figure}
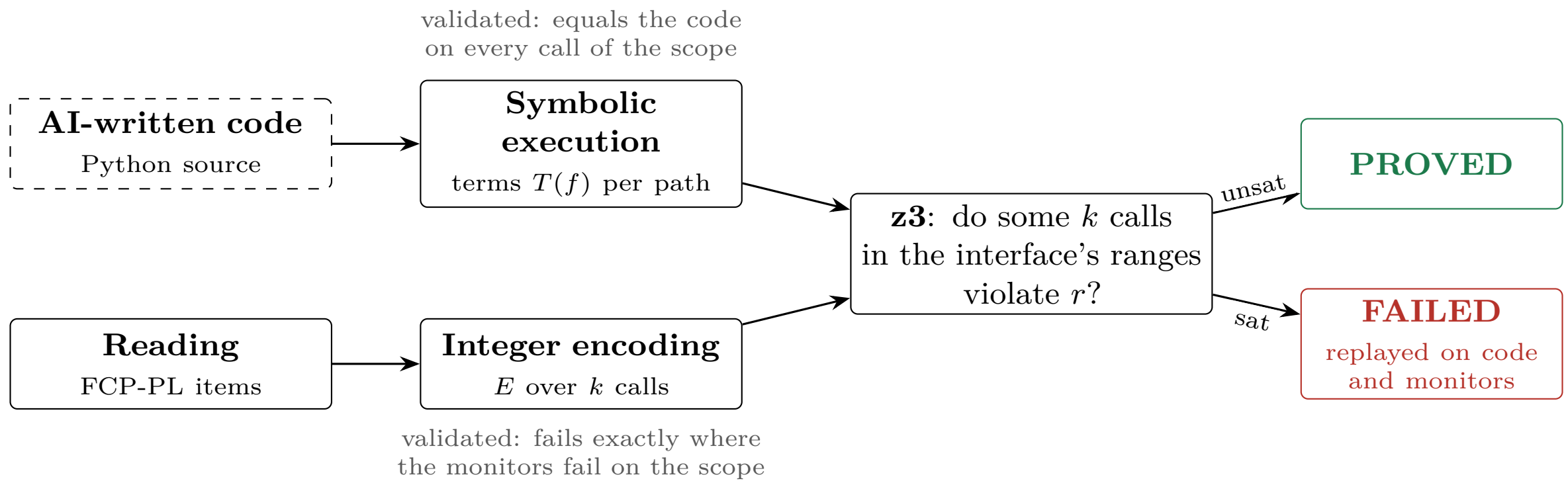

Scope evidence is exhaustive on its calls and makes no claim about other inputs. For code in a translatable subset of Python, FCP decides each requirement for every input the interface can carry (Figure~\ref{fig:prove}).

\paragraph{Symbolic execution.} FCP translates the function's source into z3 terms by symbolic execution~\cite{king1976symbolic,cadar2008klee}, following every path with its condition, so that the outcome and returned values of one call become terms over the arguments. The subset covers module constants; helper functions, inlined, including ones that raise; \code{isinstance} and \code{type()} checks, decided by the declared types; loops over \code{range()} or a literal tuple; comparisons, \code{and}/\code{or} and conditional expressions; \code{min}, \code{max}, \code{abs} and \code{divmod}; and \code{//} and \code{\%} by positive constants. FCP refuses, without approximation, floats, values whose Python type depends on the path (such as \code{x < 5 and FEE}), regular expressions, and calls to any other function.

\paragraph{Two second semantics, validated.} The translation is a second implementation of a subset of Python, so it is checked before it is trusted, in the spirit of translation validation~\cite{pnueli1998tv,necula2000tv}: on every call of the scope, its outcome and values must equal what the code returned. Compiled back to Python, the terms take about a second on the 1.36 million calls of the checkout scope. The integer encoding of the reading is a second implementation of \FCPPL{} and is checked the same way: on every cycle of the scope trace, it must report a failure exactly where the compiled monitor does. A single difference in either check blocks the proof.

\paragraph{The query.} The translation, joined to the integer encoding over $k$ calls, poses the question: do some $k$ calls, anywhere in the interface's ranges, violate $r$? If z3 answers no and the horizon is exact, $r$ is PROVED for every call the interface can carry. If z3 answers yes, its calls are run on the real code and the resulting trace is checked by the compiled monitors before the verdict FAILED is reported.

\begin{proposition}[Soundness of proofs beyond the scope]
\label{prop:prove}
(i) Every FAILED verdict is witnessed by a run of the code that the compiled monitors reject. (ii) If $T(f)$ agrees with $f$ on every input within the interface's ranges and $E$ agrees with the compiled monitors on every trace, then a PROVED verdict implies that no trace of $f$ violates $r$, and a PROVED-BOUNDED verdict implies that no sequence of at most $k$ calls does.
\end{proposition}

Part (ii) is conditional. Validation on the scope is evidence, short of a proof, that the two second semantics are faithful (\S\ref{sec:limits}). Part (ii) also takes z3's answers as given, so each formula z3 finds unsatisfiable is decided again by cvc5~\cite{barbosa2022cvc5}, which confirms all 1{,}528 proofs of Table~\ref{tab:prove} and refutes none.

\paragraph{Strings and equivalence sweeping.} In a function over strings or lists, a sequence is a list of guarded slots (a condition and a value each), so filters, loops and slices remain terms; character tests such as \code{str.isdigit} are Python's own answers on the 256 codes the interface carries. The query then sets two computations of one quantity side by side, a checksum as the code computes it and as the reading defines it, which z3 does not decide as a whole. FCP applies equivalence sweeping, the method of hardware equivalence checkers~\cite{kuehlmann2002robust}: subterms that agree on sample calls are candidate equalities, and z3 proves each from the leaves up with the proved ones substituted; a refuting model becomes a new sample. Every merge is proved, so samples affect only speed. In a goal that remains open, the small conjuncts (the facts of the path that returned) are assumed and the rest is swept again; a goal without such facts is proved outcome by outcome. The same sweep decides comparisons of readings that define one quantity in two ways.

\paragraph{Effects, files and objects.} For code with effects, each call on the service interface becomes an event in the translation, and each call that receives an answer takes the next answer of the script as a free variable; an answer delivered as an exception forks the path. Paths that made different calls stay apart, and a run lasts at most a bounded number of calls. The integer encoding gains ghost state, truncated to the ghost's width as in the compiled monitors, and the \code{next}, \code{after\_reset}, \code{once}, \code{hist} and \code{since} forms. Files the code opens are objects with their own events: \code{open} returns a handle numbered within the path, \code{with} closes it on every exit, a call on a closed file raises \code{ValueError}, a \code{try} statement resumes from the state in which the exception was raised, and each path keeps its own view of the disk, from which \code{os.path.exists} answers. For objects with several methods, the constructor yields the fields, each call runs the method that $\mathit{op}$ selects, and every way the method can end is merged into the fields after it. A call is then one step from fields and arguments to fields and results, and requirements are proved for call sequences of every length by induction~\cite{sheeran2000induction}. The state is the object's fields with the reading's ghosts and history registers; a reading that looks back $w-1$ calls is judged on a window of $w$ consecutive calls from a free state. Houdini~\cite{flanagan2001houdini} finds the strongest conjunction $I$ of template facts (comparisons among state variables and the constants of the code and the reading, also under conditions) that holds after the first calls on a new object and is preserved by every call. A requirement is PROVED if $I$ before a window implies it at the window's last call and it holds on the first $\max(1,w-1)$ calls after reset; the window encoding is proved equal to the unrolled one on the calls both cover. Otherwise FCP checks all sequences of at most $k$ calls, the scope's longest, and reports PROVED-BOUNDED. A dictionary field is translated as the sequence of its writes; in the induction it is an arbitrary function of the key and a list an arbitrary list of bounded length, with candidates first filtered on states sampled from the scope.

\section{Evaluation}
\label{sec:eval}

\subsection{Research questions and setup}
\label{sec:eval-setup}
We ask five questions.
\RQ{1}~(False assurance): how often do self-check baselines certify violated requirements, and does FCP?
\RQ{2}~(Request defects): does FCP find conflicts and under-specification in the request before any artifact exists?
\RQ{3}~(Independent readings): is disagreement between independent formalizers a reliable error signal, and when does their agreement mislead?
\RQ{4}~(Reach): do verdicts on the exhaustive scope hold for all inputs?
\RQ{5}~(Cost): what does FCP's assurance cost in questions for the author and in computation?

\begin{table}[t]
\caption{Corpus. Each request has planted defects no agent saw. Models: Claude Haiku, Sonnet and Opus~5.5.}
\label{tab:corpus}
\centering\small
\begin{tabularx}{\linewidth}{@{}l>{\raggedright\arraybackslash}p{2.9cm}>{\raggedright\arraybackslash}X>{\raggedright\arraybackslash}p{3.1cm}@{}}
\toprule
Domain & Tasks & Artifacts checked & Readings \\
\midrule
Hardware & 4 IP specifications; 13 planted defects & reference RTL, 28 single-edit RTL defects, 4 AI-generated designs & gold; 2 Sonnet per design \\
Prompt & 1 launch email, 10 instructions & 3 responses (Haiku, Sonnet, Opus); an LLM judge as baseline & gold; 2 Opus \\
Agent & 1 returns desk, 7 tools, 11 rules & 10 scaffolds; 18 runs of 3 models & gold; 2 Opus \\
Code & 10 tasks; 10 conflicts, 23 ambiguities & per task: 3 model implementations, a reference, 8 planted bugs; a refactoring in 9 tasks & gold; 2 Opus; 21 variants \\
\bottomrule
\end{tabularx}
\end{table}

\paragraph{Corpus and agents.} Table~\ref{tab:corpus} summarizes the corpus; Appendices~\ref{app:hw} to~\ref{app:code} give details (Tables~\ref{tab:hwcorpus} to~\ref{tab:codetasks}). All agents were Claude models (Haiku~5.5, Sonnet~5.5 and Opus~5.5), run between 6 and 11 November 2026 as isolated subagents, each in its own workspace. Every prompt, output and content hash is recorded, so the studies can be replayed without the models. In the hardware study, four Sonnet agents worked on each design: a decomposer, two formalizers given different style instructions, and an RTL generator. Formalizers had no access to the RTL, the gold readings or each other's work, and worked from the gold requirement list so that formalization could be measured independently of decomposition. In the other studies, the readings were written by two isolated Opus formalizers. Code writers received only the prompt, in an empty directory, and wrote an implementation and \code{pytest} tests until the tests passed.

\paragraph{Gold readings.} Gold readings are validated mechanically: they are proved on the reference designs, kill every observable mutant, and detect every planted defect. Where a request contradicts itself, the gold reading resolves it by a recorded decision. Gold readings take no part in the protocol's comparisons.

\paragraph{Metrics and baselines.} A \emph{violation} is a pair (artifact, requirement) on which the gold reading fails, with the failure replayed. A flow gives false assurance when it certifies a violation (Definition~\ref{def:fa}). The \emph{self-check baseline} certifies a requirement when the artifact passes the checks that a model wrote for it: in hardware, when the formalizer's properties for the requirement are proved on the design; in code, when the model's own test suite passes. For the prompt, the baseline is an LLM judge given a per-instruction checklist. FCP certifies a hardware requirement at level ASSURED and a code requirement as defined in Table~\ref{tab:verdicts}.

\subsection{Hardware: four designs}
\label{sec:eval-hw}

The corpus consists of a synchronous FIFO, a round-robin arbiter, a command-to-APB bridge and a two-port APB hub, with specifications written in the style of industrial IP documents. They contain planted contradictions, ambiguities, a permission that another sentence forbids, an incompleteness and a vacuity trap, and there are 28 single-edit RTL defects. About 39,700 content-addressed proof tasks were run; the median PDR obligation took 0.82\,s, and every verification obligation received a definitive verdict.

\begin{table}[t]
\caption{The hardware study (\RQ{1}, \RQ{3}). The fault operators are placeholder consequents, reset-polarity errors, masking assumptions, dropped conjuncts, same-cycle for next-cycle, and bounds off by one cycle (Appendix~\ref{app:hw}).}
\label{tab:hw}
\centering\small
\begin{tabularx}{\linewidth}{@{}>{\raggedright\arraybackslash}X>{\centering\arraybackslash}p{3.7cm}>{\centering\arraybackslash}p{3.7cm}@{}}
\toprule
& \textbf{Self-check}: AI writes assertions, the prover reports PASS & \textbf{FCP} \\
\midrule
Violated requirements certified as holding & \textbf{56 / 244 (23\%)} & \textbf{0 / 244} \\
Injected formalization faults that certify a buggy design & \textbf{71 / 161} & \textbf{0 / 161}, all 161 flagged \\
AI formalizations equivalent to intent (all well-formed and lint-clean) & 44 / 98 (45\%) & n/a (FCP does not consult gold) \\
Disagreements between the two AI formalizers that were real errors & n/a & 45 / 45 \\
Injected RTL defects detected (gold readings) & 28 / 28 & 28 / 28, all replayed \\
\bottomrule
\end{tabularx}
\end{table}

\paragraph{False assurance (\RQ{1}).} Table~\ref{tab:hw} summarizes the results. All 98 formalizations were well-formed, lint-clean and read back correctly, and 44 (45\%) were equivalent to the gold reading. The differences lay in timing windows, conditions and the resolution of conflicts. Accepting the formalizers' properties as written certified 23\% of the violated requirements, and FCP certified none. In one case, a FIFO formalizer adopted the usage rule ``the producer shall not push while full'' as an assumption. The assumption is receptive, since the environment can always refrain from pushing, so the receptiveness check correctly raises no objection. Vacuity analysis proved two of the formalizer's properties vacuous, and contract comparison showed every contract of the set to be weaker than its peer's (the other formalizer's). An earlier version of FCP that compared properties under pooled assumptions admitted this assumption into the agreement check and gave 4 false assurances on this set; comparing contracts removes them.

\begin{table}[t]
\caption{Ablation of the acceptance rule on the hardware study. Comparison in the full context raises acceptance 4.4-fold over strict equivalence, with zero false assurance under both, and weaker rules accept more, some falsely.}
\label{tab:accept}
\centering\small
\begin{tabularx}{\linewidth}{@{}>{\raggedright\arraybackslash}X>{\centering\arraybackslash}p{3.2cm}>{\centering\arraybackslash}p{3.2cm}@{}}
\toprule
Acceptance rule for a requirement & Accepted on reference RTL (readings) & False assurance on buggy RTL \\
\midrule
Self-check: the formalizer's reading is proved & 92 / 104 & 56 / 244 \\
All readings proved, with no comparison & 77 / 104 & 10 / 244 \\
Equivalent in a \emph{minimal} sufficient context, and that context proved & 64 / 104 & 5 / 244 \\
\textbf{Equivalent in the full context, and every other requirement proved under both readings} & \textbf{31 / 104} & \textbf{0 / 244} \\
Strict: equivalent without context & 7 / 104 & 0 / 244 \\
\bottomrule
\end{tabularx}
\end{table}

\paragraph{The acceptance rule (\RQ{1}, \RQ{5}).} Table~\ref{tab:accept} varies the rule by which a proved requirement is accepted without review. Strict equivalence of readings accepts 7 of 104 requirement readings on the reference designs. Equivalence in the full context, with proofs of every other requirement under both readings, accepts 31 with no false assurance. The rule is sound per requirement: the design's behaviors then lie inside the context, so the peer's reading also holds, and on every acceptance gained, the peer's reading was proved as well. All of the gain is on the bridge, the one design whose reference RTL proves every requirement under both readings; on every design, most disputes vanish in context. Requiring proof of only a minimal sufficient context accepts 64 readings and admits 5 false assurances, four of them with both formalizers wrong together.

\paragraph{AI-generated RTL and unplanted defects.} The RTL generators' designs passed their own testbenches of up to hundreds of thousands of cycles. Against the gold readings, every requirement of the generated FIFO and bridge was proved, and FCP found two deviations: one arbiter does not mask grants during reset, and one hub accepts a command in the response-handshake cycle, using a permission of the bridge that the hub's one-transaction rule forbids. FCP also found three defects that had not been planted. The hub's 12-cycle latency bound is unrealizable; the tight bound is 17 cycles. The hub's selector violates its bridge's input contract, which every end-to-end proof misses because the bridge in use samples only when idle. Finally, the gold reading of the hub omitted a requirement, which determinacy exposed.

\paragraph{Disagreement as a signal (\RQ{3}).} Across all requirements formalized by both agents, every contract disagreement between them was a real error (45 of 45). Agreement was correct 9 times and wrong twice: both arbiter formalizers read an ambiguous sentence, intended as true round robin, literally as ``shift the priority by one position''. Neither case produced false assurance, because the formalizers' properties fail on the conforming RTL and the decomposer's ambiguity flag kept the requirement in the review queue.

\subsection{A prompt}
\label{sec:eval-prompt}
A prompt for a product-launch email has ten checkable instructions. Before any response existed, the consistency analysis proved in 12\,s that no email satisfies the literal reading, and named six instructions as a minimal conflicting set: a greeting, three feature sentences each followed by a benefit sentence, and a body ``under 7 sentences''. Read literally, a greeting, three features and three benefits make seven body sentences (\RQ{2}). The readings disagree on exactly the planted ambiguities, and contextual comparison turns each disagreement into one question with an example email; in isolation, two of the four disputes arise only on emails that violate other instructions. An LLM judge (Opus), given the prompt, the three models' responses and a per-instruction checklist, marked all 33 items PASS (11 per response) and called the prompt satisfiable. It had adopted one reading (``6 body sentences, greeting not counted'') and did not report that every email fails under the other. Its reading is defensible. On the same prompt, FCP returns a question, its evidence and each answer's consequence. Two labelers (Sonnet and Haiku) differed on 1 of 189 label decisions and agreed on two that the reference labeling, written with the gold reading, rejects, one of which changes verdicts. Labeler agreement therefore does not establish correctness.

\subsection{An agent}
\label{sec:eval-agent}
The returns assistant of an online store has seven tools, a ``Talk to a human'' button and eleven rules, with one planted conflict and five planted ambiguities. Ten scaffolds were compiled from Python, each agreeing with its source on 6,000 random turns: a guarded kernel (91 lines of Python, 224 of Verilog), a plain ReAct loop~\cite{yao2023react}, and eight one-line mutants of the kernel. Before any run, forcing found in 0.18\,s that pressing the button in turn~1 makes the prompt unsatisfiable: ``hand off at once'' and ``ask whether there is anything else before ending'' cannot both hold (\RQ{2}). Under the literal reading, the guarded kernel \emph{enforces} 8 of the 9 checkable rules for every model output and violates the ninth, the conflict. On the ReAct loop every rule is \emph{delegated}, and the solver's first attack is a \$406 refund in turn~1. Six of the eight planted bugs are latent: each is unobservable as long as the model follows the rules and exploitable once it deviates, the condition that prompt injection creates (Appendix~\ref{app:agent}). Haiku, Sonnet and Opus played the assistant in six scenarios (18 runs, 82 turns). The guard blocked no proposal and forced the handoffs that the rules demand (Appendix~\ref{app:agent}). All three models ignored an injected ``pre-approved, skip confirmation'' note, and all three deviated from the literal prompt only at the planted conflict and one planted ambiguity, in the same way. Replayed through the agent's Python, the ReAct loop executes all 9 of the solver's attack traces, and the guarded kernel executes only the trace at the rule it violates by design.

\subsection{Code: a checkout function}
\label{sec:eval-code}
The checkout prompt (\S\ref{sec:intro}) has seven rules, formalized as nine requirements, including ``discounts stack, but the total discount is never more than 30\%'', ``round to the nearest cent'', free shipping from \$50, and ``adding items to the cart must never lower the amount charged''. One conflict and four ambiguities are planted. We check the three models' code, a reference and eight one-line planted bugs, each on 1,355,700 calls.

\paragraph{Before any code (\RQ{2}).} The boundary search finds the conflict under all three readings, in 17 to 65\,s: \code{checkout(4999,0,False)} must return $(4999,499)$, a charge of \$54.98, and \code{checkout(5000,0,False)} must return $(5000,0)$. Every minimal conflicting set contains the shipping and monotonicity rules. The trigger-situation check finds the readings consistent, because none of its candidates lies on the threshold. Determinacy finds one open output under every reading: at exactly half a cent, \code{checkout(5,0,True)} may cost 4 or 5 cents. The three readings differ under strict comparison, because two of them state the price rules only for calls that return. In context they are equivalent on every requirement, each decided on two calls in under a second; bit-level spot checks on 400 calls find no counterexample to any of them.

\begin{table}[t]
\caption{The checkout study (\RQ{1}). All three models stated the conflict, and each resolved it differently (Figure~\ref{fig:conflict}).}
\label{tab:code}
\centering\small
\begin{tabularx}{\linewidth}{@{}lll>{\raggedright\arraybackslash}X@{}}
\toprule
Code & Own tests & Violates (all readings) & First failing call (replayed in a fresh process) \\
\midrule
Haiku & 134 pass & R-SHIP-1 & \code{checkout(5000,0,False)} $\to$ (5000, 498) \\
Sonnet & 196 pass, 1 xfail & R-MONO-1 & \code{checkout(4999,0,False)} $\to$ (4999, 499), then \code{checkout(5000,0,False)} $\to$ (5000, 0) \\
Opus & 745 pass & R-SHIP-1 & \code{checkout(4998,0,False)} $\to$ (4998, 2) \\
\midrule
\multicolumn{4}{@{}p{\linewidth}@{}}{Planted bugs flagged in the planted requirement: 7 of 7 (C1 to C5, C7, C8). C6 (float \code{round()}, half a cent to even) is correctly not flagged: every reading allows either neighbor.} \\
\multicolumn{4}{@{}p{\linewidth}@{}}{False assurance on the models' code: the self-check baseline (own tests pass) certifies 3 of 3 violations, FCP none.} \\
\bottomrule
\end{tabularx}
\end{table}

\paragraph{The code (\RQ{1}).} Table~\ref{tab:code} and Figure~\ref{fig:conflict} show the results. All three models reported the conflict in their replies, each passed its own tests, and under every reading each violates a rule. Code splits reveal four decisions on which the implementations differ. In the largest (406 calls), Haiku and Opus measure ``an order under \$50'' on the discounted price; Sonnet, the reference and all three readings measure the subtotal. The decomposer had flagged that phrase, so it stays in the review queue independently of the formalizers' agreement. The review queue for the prompt holds three questions: which rule yields at the threshold, which amount the \$50 refers to, and how half a cent is rounded. If rule~7 yields, FCP certifies Sonnet's code on all eight remaining requirements and flags Haiku and Opus on rule~6. Until the conflict is answered, FCP certifies 9 of the 27 (requirement, implementation) pairs. The models' suites encode their authors' decisions: on the other implementations they fail up to 212 of 745 tests, and Haiku's suite misses the planted shipping bug at \$50.00, where its own code charges shipping. The only mutants surviving every reading change the rounding constant, the open half cent, or a value inside Opus's fee taper where both values violate rule~6.

\subsection{Nine more code tasks}
\label{sec:eval-beyond}
Nine further tasks, each a short prompt with one planted conflict, exercise the adapters of Section~\ref{sec:beyond}. Each was given a gold reading, read by two isolated Opus formalizers, and implemented by Haiku, Sonnet and Opus from the prompt alone (Table~\ref{tab:beyond}; Appendix~\ref{app:code}).

\begin{table}[t]
\caption{Nine more code tasks. ``Before code'': conflict search under gold, with its stimulus. ``Choice'': what all three models did at the conflict. ``Agree on'': calls or runs of the scope on which their code answers identically. ``Own tests'': suite sizes (Haiku, Sonnet, Opus). ``Bugs'': planted variants flagged in their planted requirement (a refactoring, flagged in no task, is excluded). ``Certify'': violations under gold certified by the self-check and by FCP.}
\label{tab:beyond}
\centering\footnotesize
\setlength{\tabcolsep}{3pt}
\begin{tabularx}{\linewidth}{@{}>{\raggedright\arraybackslash}p{1.4cm}>{\raggedright\arraybackslash}p{1.3cm}>{\raggedright\arraybackslash}X>{\raggedright\arraybackslash}p{1.6cm}>{\raggedright\arraybackslash}p{2.2cm}>{\raggedright\arraybackslash}p{1.4cm}>{\raggedright\arraybackslash}p{1.85cm}c>{\raggedright\arraybackslash}p{1.05cm}@{}}
\toprule
Task & Adapter & Conflict & Before code & Choice & Agree on & Own tests & Bugs & Certify \\
\midrule
Card masking & strings & last four digits vs.\ at most a quarter & 92\,s, 1 call & a quarter: 3 digits shown & every call & 38, 146, 119\textsuperscript{b} & 8/8 & 3/3, 0 \\
Rate limiter & object & under the limit vs.\ 10\,s penalty & 9\,s, 5 calls & penalty & 175,528 calls & 48, 43, 43\textsuperscript{b} & 7/8 & 3/3, 0 \\
Student pricing & change & \$75 shipping vs.\ no change for non-students & 29\,s, 1 call & \$50 for non-students & 3.6\,M calls & 22, 29, 45 & 8/8 & 3/3, 0 \\
Payment retry & effects: service & keep trying vs.\ 3 attempts & 67\,s, 3 errors & 3 attempts & 50,000 runs & 21, 36, 105 & 8/8 & 3/3, 0 \\
Atomic save & effects: files & no temporary file vs.\ keep it on failure & 18\,s, 1 failed move & keep the file & see\textsuperscript{a} & 18, 37, 33\textsuperscript{b} & 8/8 & 3/3, 0 \\
Invoice lines & records & line rounding vs.\ order rounding & 21\,s, 2 items & discount summed per line & every call & 19, 35, 23\textsuperscript{b} & 8/8 & 3/3, 0 \\
Gift card & several methods & \$5 bonus vs.\ \$500 cap & 1\,s, 2 calls & bonus counts toward the cap & 64,576 calls & 21, 38, 42 & 8/8 & 3/3, \textbf{3}\textsuperscript{c} \\
Order placement & effects: imports & undo after a failure vs.\ at most 4 calls & 46\,s, 1 refusal & undo, in 5 calls & 46,400 runs & 31, 53, 51\textsuperscript{b} & 8/8 & 3/3, 0 \\
Loyalty points & package & tier set by the points vs.\ tier never goes down & 2\,s, 4 calls & tier kept & 44,079 calls & 32, 74, 72\textsuperscript{d} & 8/8 & 6/6, 0 \\
\bottomrule
\end{tabularx}\par\vspace{2pt}
\raggedright\footnotesize\textsuperscript{a}Haiku and Sonnet make the same calls on all 1,441 runs; Opus also removes a stale temporary file first, a choice the readings leave open, so its calls differ on 1,179 runs, with every verdict in agreement. \textsuperscript{b}Each suite fails on another model's code only on behavior outside the scope (other argument types, missing keys, threads, failures the adapter does not script) or on a test of a docstring; otherwise every suite passes on all three. \textsuperscript{c}0 with a variant reading per answer of the flagged phrase (\S\ref{sec:eval-agree}). \textsuperscript{d}Each suite imports its author's internal names and does not load on another model's package.
\end{table}

\paragraph{Identical choices across models (\RQ{1}, \RQ{2}).} In each of these nine tasks, all three models made the same choice at the conflict, and in eight of them their code then answers every call or run of the scope identically. Agreement among independent implementations therefore does not indicate that a prompt is consistent. In these tasks it reflects the conflict, resolved identically three times, in line with the correlated errors observed across language models~\cite{kim2025correlated}. Every implementation violates the gold reading, the self-check baseline certifies every one of these violations, and FCP certifies none in eight of the nine tasks. Under the gold reading, FCP's search names each conflict before any code exists, with a shrunk stimulus (Table~\ref{tab:beyond}); under the formalizers' readings it names eight of the nine, the exception being the gift card's, which both formalizers' readings resolve (\S\ref{sec:eval-agree}). The trigger-situation check misses the conflict under the gold reading in all nine tasks and finds it under a formalizer's reading in two. The models themselves name most conflicts: over the ten tasks, their summaries name the conflict in 25 of 30 cases, and Opus, asked without tools, in 81 of 90 replies (model-graded, spot-checked by hand); the conflict of the loyalty package is the only one that no summary or reply names. FCP adds the conflicting calls and a verdict on the code under each resolution.

\paragraph{What the scope reaches.} Points drawn from the readings reach narrowly defined behavior: valid card numbers, the third denial inside a window, three errors in a row, a failed move after a successful write, two invoice lines whose discounts each lose a fraction of a cent, and a computed subtotal of exactly \$10.00. Under the gold reading, the scope flags 71 of the 72 planted bugs in their planted requirement. The one miss, a rate-limiter bug (a denial during the penalty does not restart it), lies at least two edits from every request-derived sequence; the models' own tests detect it, as does the proof (\S\ref{sec:eval-proofs}).

\subsection{Proofs beyond the scope}
\label{sec:eval-proofs}

\begin{table}[t]
\caption{Proofs beyond the scope (\RQ{4}). A pair is an implementation (the three models, the reference and the planted variants) under one reading. A failing proof carries inputs chosen by z3, replayed on the code; ``Missed'': requirements that pass the scope and fail the proof. The scope is recomputed for the proofs. $^\dagger$PROVED-BOUNDED: no sequence of up to 10 calls (gift card) or 8 to 15 calls (rate limiter) fails; the four on the gift card fail on longer ones. The other proofs on these two objects cover call sequences of every length.}
\label{tab:prove}
\centering\small
\begin{tabular*}{\linewidth}{@{\extracolsep{\fill}}lrrrrrrr@{}}
\toprule
Task & Scope & Pairs translated & Verdicts & Proved & Fail in both & Missed & s per pair \\
\midrule
Checkout & 1,355,700 calls & 30 / 36 & 260 & 198 & 62 & 0 & 6 to 7 \\
Card masking & 131,613 calls & 36 / 39 & 240 & 161 & 75 & 4 & 6 to 97 \\
Rate limiter & 175,528 calls & 39 / 39 & 208 & 58\,+\,7\textsuperscript{$\dagger$} & 140 & 3 & 28 to 734 \\
Student pricing & 3,601,728 calls & 36 / 39 & 204 & 139 & 65 & 0 & 24 to 27 \\
Payment retry & 50,000 runs & 39 / 39 & 364 & 293 & 71 & 0 & 1 to 2 \\
Atomic save & 1,441 runs & 36 / 39 & 276 & 208 & 68 & 0 & 0.3 to 0.5 \\
Gift card & 65,771 calls & 48 / 52 & 360 & 182\,+\,4\textsuperscript{$\dagger$} & 174 & 0 & 0.6 to 57 \\
Order placement & 46,400 runs & 39 / 39 & 351 & 278 & 73 & 0 & 1 to 2 \\
\midrule
Total & & 303 / 322 & 2,263 & 1,517\,+\,11\textsuperscript{$\dagger$} & 728 & 7 & \\
\bottomrule
\end{tabular*}
\end{table}

Table~\ref{tab:prove} reports the proofs on the eight tasks whose code falls in the translatable subset (\RQ{4}). Of 322 pairs of an implementation and a reading, 303 translate; the other 19 involve planted variants that call \code{round()} on a float, divide as floats or use a regular expression, which the translator refuses by design. The 728 requirements the scope failed also fail the proof; of the 1,535 it passed, 1,517 are proved, 11 are proved up to a bound and 7 fail. A proof extends a verdict from millions of scope calls to every input and, on the two objects, to every sequence of calls. Four of the failures the scope missed are on card masking, at strings its 131,613 calls do not hold: a variant that strips the number first accepts one that begins with a carriage return or a no-break space, under all three readings, and a variant that tests digits with \code{str.isdigit} accepts superscript digits and then shows half of the digits 0 to 9. The other three are on the rate limiter; one is the planted bug the scope missed, found at six calls, so all 72 planted bugs are flagged. Four bounded verdicts are false. They concern three gift-card variants whose balance drifts, under the gold reading, whose model of the balance is a ghost with a declared range: a long sequence takes the model out of range, and the requirement fails on the compiled monitors at calls 21, 41 and 4,155, far beyond the 10 calls of the bounded proof. The induction found no invariant because none exists. The other seven concern two limiter variants whose lists the invariant search cannot bound (one grows with every denial, one is shared by all customers); they are not known to fail. PROVED-BOUNDED marks exactly the verdicts that length can still overturn. On the change task, ``nothing changes for non-students'' is proved for all three models' code under every reading, over every call the interface carries. On the payment retry, every model's code is proved on every requirement other than the conflicting rule, for every amount and every script of service answers. On the gift card, the models' code is proved on every requirement under both formalizers' readings: the shared misreading (\S\ref{sec:eval-agree}) holds beyond the scope too, so proofs strengthen conformance and leave fidelity unchanged.

\subsection{When independent formalizers agree and are wrong}
\label{sec:eval-agree}
The gift-card prompt requires a ValueError ``for a load that would take it over'' the \$500 cap and leaves open whether the \$5 bonus for a load of \$100 or more counts. Read literally (gold), a load of \$495.01 must be accepted, and its bonus violates the cap. Both formalizers and all three models counted the bonus toward the cap, which removes the conflict and rejects that load. The decomposer had flagged the phrase on the cap requirement, which stayed a question. Its answer also decides the load requirement, which carries no flag, so FCP certified the three violations, the only false assurance in the code studies (\RQ{3}).

We therefore added a \emph{variant reading}: one more isolated formalizer, instructed to adopt the reading of the flagged phrase that no formalizer had chosen. Every requirement on which the variant differs from a formalizer's reading in context depends on the answer and goes to the author. This flags exactly the cap and load requirements, and FCP then certifies none of the violations (18 of 24 pairs, down from 21). We designed this step after observing the failure, then ran it on all ten tasks, where variant $k$ takes the $k$-th listed answer of every flagged phrase (20 more formalizers, \$25). It removes the three false certificates and holds back 12 of the 87 true ones: a variant that assumes more than the formalizers differs from both on nearly every requirement. Holding such a requirement only where a variant may reject a run that a formalizer accepts, strictly or in context, a rule written after seeing these results, removes the same three and holds back 3. The automated pipeline (\S\ref{sec:eval-cost}) applies the first rule before any code and this rule given the code. Given the code, it asks of each requirement on which a variant differs in context whether the variant accepts every run that both formalizers accept, with and without context, and releases the requirement only when both searches close in that direction. On the stored relations of the 21 variants this holds 39 requirements, where the designed rule holds 48 and the second rule 27, and it leaves 84 of the 87 true certificates and no false one. Two adversarial reviews of the engine shaped this rule: the first showed that closure in context alone is unsound, since a run may meet both formalizers and break another rule of the variant, and the second found no false release in the corrected rule.

The atomic-save task shows the same mechanism without false assurance. Both formalizers read the rule on \code{os.fsync} as not requiring a flush of Python's buffer first, a reading the prompt's example supports and the gold reading rejects. A planted bug that calls \code{os.fsync} without flushing, a real durability defect that the gold reading flags, therefore passes every test of every model and every requirement of both formalizers' readings. In all, independent formalizers agreed against the gold reading in three tasks (the arbiter, the gift card and the atomic save), each time on a phrase flagged as ambiguous. A flag on one requirement carries no information about the other requirements that its answer decides.

\subsection{End-to-end automation and cost}
\label{sec:eval-cost}
Every study above was assembled by hand. The command \code{fcp review} performs the same steps from the request and the code alone: an agent writes the adapter from the request and the code's public signature (its body is withheld); a decomposer writes anchored requirements; two isolated formalizers, and a variant formalizer where needed, write readings; and FCP runs the analyses and reports certified requirements and questions. On a duration parser written by Sonnet (12 minutes of agents, 31 of analyses), the report certifies 5 requirements and lists 13 open items, among them a replayed failure (\code{parse\_duration("002h")} raises) and the request's own contradiction (parts up to 99 allow ``99h'', which ``a duration is never longer than one day'' excludes). On the gift-card prompt with Opus's code, the report certifies 7 requirements, none of them violated under the literal reading, and asks 24 questions; the study's false assurance does not recur.

To test FCP on requests we did not write, we ran \code{fcp review}, with a signature and no code, on 20 HumanEvalComm problems~\cite{wu2025humanevalcomm}: each original description and its ambiguous, inconsistent, and incomplete rewrites (80 requests, one run). FCP completed a review for 72; the other eight exceeded a time or memory limit, seven of them on arithmetic that FCP-PL lacks. A question backed by a solver result was asked on 34 of 53 reviewed rewrites (read by hand, it concerns the rewrite's change on 33; on 27 it shows a concrete call) and on 3 of 19 reviewed originals. The same model without tools, instructed to ask before coding, asks about 56 of 60 rewrites and 1 of 20 originals. On such short single-function requests, a review therefore adds a solver result behind a question and no additional recall.

Avoiding false assurance requires review (\RQ{5}). On the reference hardware designs, the queue holds 8 to 27 decisions per reading, and with Sonnet formalizers 7 to 31 of 104 requirement readings were accepted without review, depending on the rule (Table~\ref{tab:accept}). In the code tasks (requirements times three implementations), FCP certifies 9 of 27 pairs on the checkout, 9 of 21 on card masking, 12 of 18 on student pricing, 9 of 27 on the payment retry, 6 of 24 on the atomic save and on invoice lines, 18 of 24 on the gift card with the variant reading, and none on the rate limiter, where every requirement fails under a formalizer's reading or is read differently by the two. Because the questions concern the request, one answer settles them for every implementation. A hardware study run takes 6 to 26 minutes per design, a code study minutes to hours, and a proof beyond the scope seconds per pair, and longer for long object sequences.

\section{Discussion}
\label{sec:discussion}

\paragraph{The evidential value of agreement.} Every form of agreement we examined failed at least once: code and its own tests (all 30 implementations), three models' code agreeing on every call (eight tasks), two formalizers (three tasks), two labelers, and an LLM judge that passed every response to an unsatisfiable prompt. Approaches that select or validate outputs by agreement, between programs and generated tests~\cite{chen2023codet} or among sampled answers~\cite{wang2023selfconsistency}, inherit this limit whenever the agreeing parties share a misreading. Disagreement was reliable where we measured it: all 45 contract disagreements between hardware formalizers were real errors. FCP is built around this asymmetry: it turns disagreements into witnesses, keeps every flagged phrase open irrespective of agreement on it, and asks the author whenever no machine can decide.

\paragraph{What FCP trusts.} FCP's verdicts rest on four components: the compiler from \FCPPL{} to monitors, checked by two engines, by replay, and against a reference semantics in Lean (Appendix~\ref{app:proofs}); the engines, checked against each other; the adapters, whose frame rules enter every reading and whose label semantics the analyses exercise; and the second semantics for comparison and proofs, validated on every call of the scope. Language models write requirements, readings, labels and adapters, and every verdict that depends on a model's judgment is backed by a mechanical check or a question to the author.

\paragraph{Questions about the request.} Because FCP's questions concern the request, they are asked once and reused. An answer becomes a recorded decision that every later reading must honor, and every artifact written for the request is then judged against it. Related interactive approaches ask the user to approve tests of one implementation~\cite{lahiri2022ticoder} or to clarify a prompt before generation~\cite{mu2023clarifygpt}. FCP derives each question from the readings and attaches the behavior that its answer decides.

\paragraph{Toward a confidence layer over AI assistants.} The end-to-end pipeline suggests a deployment in which every substantial answer of a coding assistant arrives with a report: the requirements that hold, with their evidence, and the open questions, each with a concrete call. Work on scalable oversight asks how people can supervise outputs they cannot fully evaluate~\cite{amodei2016concrete,irving2018debate,bowman2022oversight}; FCP contributes one mechanism, which reduces the author's task from judging an artifact to answering concrete questions about the request.

\section{Related Work}

\paragraph{Formal analyses.} Vacuity~\cite{beer1997vacuity,kupferman2003vacuity}, coverage metrics for formal verification~\cite{chockler2003coverage}, assume-guarantee reasoning~\cite{mcmillan1999circular,benveniste2018contracts}, realizability~\cite{pnueli1989synthesis}, self-composition and hyperproperties~\cite{barthe2004selfcomposition,clarkson2010hyperproperties}, bounded model checking~\cite{biere1999bmc,clarke2004cbmc}, induction-based safety checking~\cite{sheeran2000induction} and IC3/PDR~\cite{bradley2011ic3,een2011pdr} supply individual analyses. FCP assigns each to a situation of Table~\ref{tab:modes} and adds refinements that our studies required: assert and assume forms of every property, feasibility-based conflicts, extendable witnesses, and contract-level comparison in context. Translation validation~\cite{pnueli1998tv,necula2000tv} inspires our second-semantics checks, and proof-carrying code~\cite{necula1997pcc} the report.

\paragraph{Language models for specifications and verification.} Language models translate natural language into temporal logic~\cite{cosler2023nl2spec}, generate hardware assertions~\cite{kande2023assertions,orenesvera2023llmfv}, write program postconditions~\cite{endres2024nl2postcond} and specifications~\cite{ma2025specgen,wen2024autospec}, and generate verified code in Dafny~\cite{misu2024dafny,sun2024clover}. They also write RTL~\cite{thakur2023verilog,liu2023verilogeval} and code~\cite{chen2021codex,jimenez2024swebench}. These works ask whether a model can produce a specification or a verified artifact (Clover, for example, checks consistency among code, annotations and docstrings). FCP addresses the complementary question: whether a consistent and verified specification means what the request means. Our hardware study quantifies the gap (45\% of well-formed formalizations matched the gold reading).

\paragraph{Testing AI-written code and intent.} Test suites under-approximate correctness~\cite{liu2023evalplus}, and deciding what an output should be is the oracle problem~\cite{barr2015oracle}. Generated tests can rank generated programs~\cite{chen2023codet}, and interactive approaches formalize intent through tests approved by the user~\cite{lahiri2022ticoder} or through clarifying questions~\cite{mu2023clarifygpt}. Ambiguity in requirements is a long-studied source of defects~\cite{berry2003ambiguity}. In FCP, the oracle is a pair of independent readings of the request, questions carry witnesses computed from them, and the evidence on code is exhaustive within a request-derived scope~\cite{jackson2006alloy} or a proof beyond it~\cite{king1976symbolic,cadar2008klee}, complementing random property-based testing~\cite{claessen2000quickcheck}; mutation~\cite{jia2011mutation} and differential testing~\cite{mckeeman1998differential} supply two of its checks.

\paragraph{Judges, correlated errors and oversight.} LLM judges~\cite{zheng2023judge} return a single verdict, can be swayed by adversarial content~\cite{raina2024judgerobust}, and favor their own generations~\cite{panickssery2024selfpref}; self-correction without external feedback is also unreliable~\cite{huang2024selfcorrect}. Errors of different language models are correlated~\cite{kim2025correlated}, echoing the finding that independently developed program versions fail together~\cite{knight1986independence,avizienis1985nversion}; we observe the same effect for formalizers, labelers and code writers, and design the protocol around it.

\paragraph{Agents and prompt injection.} Indirect prompt injection~\cite{greshake2023injection} motivates the cut point. Benchmarks measure attacks and defenses~\cite{debenedetti2024agentdojo} and tool-agent-user interaction~\cite{yao2024taubench}, and system designs confine the model by construction~\cite{debenedetti2025camel}. Runtime verification~\cite{leucker2009rv} and shield synthesis~\cite{bloem2015shield} are the closest formal relatives. FCP complements them by proving, for a given scaffold and prompt, which rules hold for every model output and which rest on the model, before the agent runs.

\section{Limitations}
\label{sec:limits}

\paragraph{Scale and corpus.} The hardware designs have a few hundred state bits, and the corpus contains one prompt, one agent and ten code tasks, nine of one function or class and one a package of three modules. The planted defects were chosen by us, and real requests are longer, with defects that may be subtler.

\paragraph{Gold readings and models.} The gold readings were written with a model of the formalizers' family. They are validated mechanically (proved on reference designs, killing every observable mutant, detecting every planted defect), and their fidelity to the prose remains a judgment, recorded as decisions. All formalizers, code writers and judges are Claude models, and other providers' models may err differently.

\paragraph{Correlated formalizers.} Formalizers agreed against the gold reading on flagged phrases in three tasks, and in the gift card FCP certified three violations. A variant reading per answer removes them at two or three more formalizers per task. Its coverage extends only to questions that were asked: on 8 requirements both formalizers accept a run that the gold reading rejects, and a variant can reject such a run on 5.

\paragraph{Exactness of the adapters.} Labels bound the exactness of text verdicts, and tone or helpfulness remain judged by people. For agents, delegated rules are measured by runs and remain unproved, and the Python-to-Verilog translation is checked by co-simulation and is likewise unproved. For code, the evidence is exhaustive within a scope; it is a proof only where the code translates to z3 terms (functions on integers, strings and lists, changes to them, calls on a service, files, and call sequences on objects whose fields are numbers, dictionaries or lists, for every length only where an invariant is found), and in that case it is conditional on the second semantics (Proposition~\ref{prop:prove}). A file's contents are not modeled, and a package of several modules is checked on its scope without a proof. Comparison witnesses for objects and changes are bounded counterexamples, which need not be the simplest calls, and a look-back inside \emph{once} delays the cover that contains it.

\paragraph{Bounded searches and cost.} Forcing, boundary conflicts and fixed-stimulus triage are sound for what they report and incomplete; a clairvoyant implementation can meet a bound on any single stimulus, so the hub's unrealizable bound was confirmed by a sweep and by hand. Low false assurance is obtained at the cost of review questions, and review effort is reported as a count of decisions; human time was not measured. Finally, formalizers were isolated by instruction and by workspace; no sandbox was used.

\section{Conclusion}

When the same kind of system writes an artifact and its checks, passing checks carry little evidence. Across four kinds of artifact, FCP found what the producing models' own checks missed: contradictions in the requests, before any artifact existed; readings that differ from the intent, each shown by one concrete example; rules that rest on a model's compliance; and decisions that independent implementations made differently or, more often, identically. Of the violated requirements it was shown, FCP certified three, all in one code task where both formalizers and all three models made the same choice on a flagged phrase; a variant reading removes them. Agreement between independent formalizers bounds how much is assured without review, and each disagreement is concrete because FCP turns it into a waveform, a conversation or a call. Future work includes testing the second variant rule on new tasks, larger requests and programs, and proofs for packages of several modules.

\section*{Broader Impact}
FCP is intended to reduce the risk that unreviewed AI-generated designs, agents and code are deployed on the strength of self-written checks. Its main risk is misplaced confidence in its verdicts: a certified requirement inherits every error that its agreeing readings share, and our gift-card study shows that such errors occur. FCP's reports therefore state the evidence behind each verdict, keep every flagged phrase in the queue, and present each proof as a statement about a reading, leaving intent to the author. The agent adapter makes explicit which rules an agent's scaffold enforces and which rest on its model; we see this as a tool for defenders, and the solver's counterexamples are also attack traces against unguarded scaffolds.

\section*{Reproducibility}
The repository, \url{https://github.com/shankch/FCP_oss}, holds the prototype, corpora, agent prompts and transcripts, and all results. Every agent output is content-addressed and recorded, so every study replays without access to the models. Each study is regenerated by one script under \code{experiments/}, which writes the JSON results that the tables report, with tool versions. The prototype's 228 tests check the adapters, the translator and the encodings.

\bibliographystyle{plainnat}
\bibliography{refs}

\appendix

\section{Proofs}
\label{app:proofs}

\paragraph{Machine-checked statements.} Propositions~\ref{prop:conflict} to~\ref{prop:prove} and Corollary~\ref{cor:pin} are proved in Lean~4~\cite{demoura2021lean4} (84 theorems, standard axioms only), with two facts they use: every \FCPPL{} item is causal, and each register the compiler emits (every operator; every pattern other than \code{inorder}) computes an executable reference semantics in closed form. On random traces of 46 readings, simulated monitors agree with it on all 1{,}747{,}852 compared cycles, after it exposed three compiler defects outside the corpus. Lean checks the registers as transcribed; the comparison tests the Verilog, noticing 1{,}436 of 1{,}840 sampled changes to it (14 of 120 for string readings).

\paragraph{Proposition~\ref{prop:conflict}.} Let $M$ accept $\sigma$, that is, let some $\pi\in\sem{M}$ carry the inputs $\sigma$ in cycles $0,\dots,k$, and let $\tau$ be the outputs of $\pi$ in those cycles. By Definition~\ref{def:conflict}, applied to $\tau$, there is a cycle $t\le k$ at which some guarantee $\varphi$ fails on $(\sigma,\tau)$, with every assumption holding up to $t$. Since monitors are causal, the verdicts on $\pi$ up to cycle $k$ are those on $(\sigma,\tau)$, so $\pi$ witnesses $M\not\models(A\Rightarrow\varphi)$ by Definition~\ref{def:ag}. \qed

\begin{corollary}[Pinned instances]
\label{cor:pin}
Let a guarantee $\forall u\,\varphi(u)$ range over a rigid variable $u$, and let $u_1,\dots,u_m$ be values. If $\sigma$ is a conflict of the readings in which $\forall u\,\varphi(u)$ is replaced by $\varphi(u_1)\wedge\dots\wedge\varphi(u_m)$, then $\sigma$ is a conflict of the original readings.
\end{corollary}
\emph{Proof.} Where an instance $\varphi(u_i)$ fails, so does $\forall u\,\varphi(u)$, and the assumptions are unchanged; so outputs that violate the pinned set at $t$ with the assumptions held up to $t$ violate the original set at $t$. \qed

\paragraph{Proposition~\ref{prop:trust}.} The model is a function from the history of a conversation to a proposal, and every such function produces, together with any environment, a sequence of values of $X$. Because $X$ is a free input of $M$, every run of the agent is a trace of $M$. If $M\models(A\Rightarrow\varphi_r)$, every run satisfies $r$ under $A$. If $M\models(A\wedge\hat C_r\Rightarrow\varphi_r)$, then by Definition~\ref{def:ag}, $\varphi_r$ holds at every turn $t$ up to which both $A$ and $\hat C_r$ have held; this covers every turn up to the first in which the model's proposal violates $\hat C_r$. \qed

\paragraph{Proposition~\ref{prop:horizon}.} Write cycle $0$ for reset and cycles $1,2,\dots$ for calls. Under the hypotheses, the verdict of an item at a call $t$ is a function of the calls $t-\delta,\dots,t$, where $\delta\le d$ is the item's look-back, if $t-\delta\ge 1$; otherwise the look-back reaches before reset, and the item is disabled and holds. Without assumptions, the contract $C^r_x$ is the set of traces on which every item of $r$ under $x$ holds at every call. Readings without future obligations need no extension of witnesses, since every violation-free prefix has a conforming continuation.

Suppose $\pi\in (C^r_a\cap K)\setminus C^r_b$, where $K$ is the set of all traces for strict comparison and $K^r$ for comparison in context. Some item of $r$ under $b$ fails at a call $t$ of $\pi$. Let $s=\max(1,t-d)$, and let $\pi'$ consist of reset followed by the calls $s,\dots,t$ of $\pi$: at most $d+1$ calls. Because the function has no state, the answer to each call depends only on its arguments, so $\pi'$ is a trace of the function, and trivially a trace of $\mathcal{S}_\Sigma$. Map call $j$ of $\pi'$ to call $s+j-1$ of $\pi$. The failing item sees the same window at the last call of $\pi'$ as at $t$ in $\pi$ (if $s=1$, $\pi'$ is a prefix of $\pi$), so it fails, and $\pi'\notin C^r_b$. Consider any other item $\iota$, of $r$ under $a$ or of any $r'\neq r$ under $a$ or $b$, at a call $j$ of $\pi'$. If its look-back fits in $\pi'$, its window equals its window at $s+j-1$ in $\pi$, so its verdict is the same, and it holds there because $\pi\in C^r_a\cap K$. Otherwise its look-back reaches before reset in $\pi'$, and it is disabled and holds. Hence $\pi'\in C^r_a\cap K$. The converse direction is immediate. \qed

\paragraph{Proposition~\ref{prop:prove}.} Part (i) holds by construction: a FAILED verdict is reported only after z3's calls (or runs) have been executed on the code and the resulting trace has been rejected by the compiled monitors. For part (ii), suppose some trace of $f$ violates an item of $r$ at call $t$. The window argument above, applied to a single item, shows that the at most $k$ calls ending at $t$, run from reset, also violate it. (The horizon is exact when the hypotheses of Proposition~\ref{prop:horizon} hold for the items of $r$; for code with effects, $k$ covers a complete run; for objects, a PROVED-BOUNDED verdict claims only sequences of at most $k$ calls.) Because $T(f)$ agrees with $f$ on these inputs, the translated outputs equal the code's; because $E$ agrees with the monitors, the encoding fails at the same call. The query is therefore satisfiable, which contradicts the PROVED verdict. For objects, $I$ holds in the first $w$ states after reset and, if in $w$ consecutive states, in the next, hence in every state of every trace. A violation at a call $t\ge w$ would be a window that starts where $I$ holds and violates $r$ at its last call, and one at $t<w$ lies in the unrolled prefix; the queries exclude both. \qed

\section{The Property Language and an Example Reading}
\label{app:lang}

\begin{table}[!htb]
\caption{\FCPPL{} in brief. Every property is compiled to an assert form and a deterministic assume form.}
\label{tab:lang}
\centering\small
\begin{tabularx}{\linewidth}{@{}l>{\raggedright\arraybackslash}X@{}}
\toprule
Construct & Meaning \\
\midrule
Types & Booleans; bit-vectors; integers with statically computed ranges; bounded sequences (strings, lists) of at most $n$ slots \\
Expressions & exact arithmetic, comparisons, Boolean connectives on one-bit operands, conditionals, $\mathit{len}(s)$, $s[i]$, $x\in s$, ASCII character tests, $\mathit{all}/\mathit{any}/\mathit{count}/\mathit{sum}$ over $i<b$ with $\bar b\le 256$ \\
Past time & $\mathit{prev}(e,k)$, disabled before the current reset segment; $\mathit{once}$, $\mathit{hist}$, $\mathit{since}$ \\
Patterns & \code{invariant}, \code{implies}, \code{next}, \code{within}, \code{hold\_until}, \code{bounded\_wait}, \code{inorder}, \code{after\_reset}; each monitor flags the minimal bad prefix \\
State and quantifiers & ghost variables with reset values and update rules; rigid variables, universally quantified over a range; named definitions \\
Item roles & property (guarantee), assumption, scenario (cover), don't-care; each item names the requirement it formalizes \\
\bottomrule
\end{tabularx}
\end{table}

Table~\ref{tab:lang} summarizes \FCPPL{}. The listing below gives the checkout task's gold reading of the shipping and monotonicity rules. Each requirement carries its verbatim quote, line span and any recorded ambiguity (here, ``Orders under \$50'').

\begin{verbatim}
- id: G-SHIP
  req: R-SHIP-1
  pattern: implies
  args: {trigger: valid, cond: 'ship == (subtotal < 5000 ? 499 : 0)'}
  note: 'decision: the order value is the subtotal before discounts'
- id: G-MONO
  req: R-MONO-1
  pattern: implies
  args:
    trigger: 'valid && prev(valid) && outcome == 0 && prev(outcome) == 0
              && coupon == prev(coupon) && member == prev(member)
              && subtotal > prev(subtotal)'
    cond: 'price + ship >= prev(price + ship)'
\end{verbatim}

The look-back of \code{G-MONO} is one call, so Proposition~\ref{prop:horizon} decides comparisons of these readings on two calls, and the conflict of Figure~\ref{fig:conflict} is a pair of calls on which no outputs satisfy both items.

\section{Hardware Study Details}
\label{app:hw}

\begin{table}[!htb]
\caption{Hardware corpus. ``RTL lines'': the reference design; ``Request defects'': planted defects of the specification. Decomposers recovered every gold requirement (by quote overlap) and flagged every planted defect and three real ones that had not been planted. They flagged 73 to 95\% of their requirements as ambiguous: high recall, low precision.}
\label{tab:hwcorpus}
\centering\small
\begin{tabularx}{\linewidth}{@{}Xrrrrr@{}}
\toprule
Design & Spec lines & RTL lines & Gold reqs. & RTL defects & Request defects \\
\midrule
SFIFO-01 synchronous FIFO (depth 5) & 86 & 53 & 21 & 8 & 4 \\
RRA-4 round-robin arbiter & 53 & 44 & 12 & 7 & 4 \\
BRG-1 command-to-APB bridge & 84 & 61 & 21 & 8 & 3 \\
HUB-2 two-port APB hub (with BRG-1) & 45 & 124 & 13 & 5 & 2 \\
\bottomrule
\end{tabularx}
\end{table}

\begin{table}[!htb]
\caption{Injected formalization faults on the gold readings (11 of 172 injected faults were equivalent to the original and excluded automatically). ``Self-check FA'': faults that would certify a buggy design.}
\label{tab:faults}
\centering\small
\begin{tabular}{@{}lrrrl@{}}
\toprule
Fault operator & Faults & Self-check FA & FCP FA & Flagged by \\
\midrule
placeholder consequent (\code{\dots\ || true}) & 50 & 30 & 0 & vacuity or tautology, peer \\
reset-polarity error in the trigger & 27 & 18 & 0 & vacuity, peer \\
masking assumption (\code{assume !trigger}) & 27 & 20 & 0 & vacuity, peer, receptiveness \\
dropped conjunct & 17 & 2 & 0 & peer, vacuity \\
next cycle replaced by same cycle & 25 & 1 & 0 & failure on reference, peer \\
bound one cycle too loose & 11 & 0 & 0 & peer \\
bound one cycle too tight & 4 & 0 & 0 & failure on reference, peer \\
\midrule
total & 161 & 71 & 0 & all 161 flagged \\
\bottomrule
\end{tabular}
\end{table}

\paragraph{Per-set results.} The baseline's 56 false assurances fall in seven of the eight formalizer sets, from 2 of 20 violations (arbiter, set B) to 30 of 53 (FIFO, set A). Under the gold environment, 44 of the formalizers' 98 properties are equivalent to gold; the others are stronger, weaker, incomparable or, once, missing.

\section{Agent Study Details}
\label{app:agent}

\begin{table}[!htb]
\caption{Trust classes under the gold reading (E enforced, D delegated, B broken); the columns are the rules R-ID-1 to R-END-1 and the frame. Six mutants (G1 to G5, G7) are latent: each turns an enforced rule into a delegated one, which is unobservable as long as the model follows the rule. Each scaffold verifies in under a second.}
\label{tab:trust}
\centering\footnotesize
\setlength{\tabcolsep}{3pt}
\begin{tabular}{@{}lcccccccccc@{}}
\toprule
Scaffold & ID1 & ID2 & ELIG & CONF & TOT & BIG & ONCE & HUM & END & frame \\
\midrule
guarded kernel & E & E & E & E & E & E & E & E & \textbf{B} & ok \\
ReAct loop & D & D & D & D & D & D & D & D & D & ok \\
G1 confirmation not tied to the order & E & E & E & \textbf{D} & E & E & E & E & B & ok \\
G2 amount not checked against it & E & E & E & \textbf{D} & E & E & E & E & B & ok \\
G3 total from any order's lookup & E & E & E & E & \textbf{D} & E & E & E & B & ok \\
G4 handoff after the third failure & E & \textbf{D} & E & E & E & E & E & E & B & ok \\
G5 button needs an empty proposal & E & E & E & E & E & E & E & \textbf{D} & B & ok \\
G6 handoff needs verification & E & \textbf{B} & E & E & E & E & E & E & B & ok \\
G7 eligibility never revoked & E & E & \textbf{D} & E & E & E & E & E & B & ok \\
G8 handoff does not end the loop & E & E & E & E & E & E & E & E & B & \textbf{broken} \\
\bottomrule
\end{tabular}
\end{table}

\paragraph{Readings and runs.} The formalizers' readings differ from gold on two rules: whether a later ``No'' revokes an earlier ``Yes'' to a refund, and whether blocking a refund over \$150 suffices or a handoff is also required. Under their reading of the latter, the rule is delegated on the guarded kernel, so the trust class depends on the author's answer. In the 18 runs, the guard mattered only when the customer failed verification twice: all three models explained the failure and proposed no handoff, and the scaffold forced the handoff.

\section{The Ten Code Tasks}
\label{app:code}

\begin{table}[!htb]
\caption{The code tasks, each with a reference and eight planted bugs; the later nine also have a refactoring.}
\label{tab:codetasks}
\centering\small
\begin{tabularx}{\linewidth}{@{}lc>{\raggedright\arraybackslash}X@{}}
\toprule
Task & Rules & Planted conflict (number of planted ambiguities) \\
\midrule
Checkout & 7 & free shipping from \$50 vs.\ the charge never drops as items are added (4) \\
Card masking & 4 & keep the last four digits vs.\ never show more than a quarter, for 12 to 15 digits (2) \\
Rate limiter & 5 & allow below 3 attempts in 60\,s vs.\ deny for 10\,s after a denial (3) \\
Student pricing & 4 & free shipping from \$75 vs.\ nothing changes for non-students (2) \\
Payment retry & 7 & keep trying until the charge succeeds vs.\ at most 3 attempts (2) \\
Atomic save & 7 & leave no temporary file vs.\ keep it if \code{os.replace} fails (2) \\
Invoice lines & 6 & discount rounded per line vs.\ per order (2) \\
Gift card & 6 & \$5 bonus on a load of \$100 or more vs.\ \$500 cap, read literally (2) \\
Order placement & 8 & refund and release after a failed booking vs.\ at most four calls on the services (2) \\
Loyalty points & 9 & the tier follows the points vs.\ the tier never goes down (2) \\
\bottomrule
\end{tabularx}
\end{table}

\paragraph{Per-task observations.} \emph{Card masking:} the conflict is \code{mask\_card("219011401821")}, twelve digits that pass the Luhn check; the readings define the Luhn sum differently, and only sweeping decides their comparison. \emph{Atomic save:} the first conflict search exposed an escape in the gold reading, now closed. \emph{Order placement:} the conflict is \code{place\_order(7, 3, 1, 1)} with the booking refused. \emph{Loyalty points:} the conflict is \code{earn(50000)}, \code{tier()}, \code{redeem(1)}, \code{tier()}; each model's suite kills one mutant of its own package that the scope misses.

\end{document}